\documentclass[]{jfm}

\usepackage{graphicx}
\usepackage{newtxtext}
\usepackage{newtxmath}
\usepackage{natbib}
\usepackage{hyperref}
\usepackage{amssymb}
\usepackage{arydshln}
\usepackage{comment}
\hypersetup{
    colorlinks = true,
    urlcolor   = blue,
    citecolor  = black,
}
\renewcommand{\d}[0]{\textnormal d}

\title{Structure and scaling of inclined gravity currents}

\author{Lianzheng Cui
  \corresp{\email{lianzheng.cui22@imperial.ac.uk}},
  Graham O. Hughes
 \and Maarten van Reeuwijk}

\affiliation{Department of Civil and Environmental Engineering, Imperial College London, London SW7 2AZ,UK}

\begin{document}
\maketitle
\begin{abstract}
We explore the fundamental flow structure of inclined gravity currents with direct numerical simulations. A velocity maximum naturally divides the current into inner and outer shear layers, which are weakly coupled by exchange of momentum and buoyancy on timescales that are much longer than the typical timescale characterizing either layer. The outer layer evolves to a `self-similar' regime with flow parameters taking constant characteristic values. The flow behaviour in the outer layer is consistent with that found in a current on a free-slip slope by van Reeuwijk et al. (\textit{J. Fluid Mech.}, vol. 873, 2019, pp. 786–815), and the integral buoyancy forcing in the layer is balanced solely by  entrainment drag. The inner layer evolves to a quasi-steady state, in which the buoyancy forcing is approximately balanced by wall drag. The inner layer can be further decomposed into viscous and turbulent wall regions that have much in common with fully developed open channel flow. Using scaling laws within each layer and a matching condition at the velocity maximum, we solve the entire flow system as a function of slope angle $\alpha$, in good agreement with the simulation data. We further derive an entrainment law from the solution, which exhibits relatively high accuracy across a wide range of Richardson numbers and provides new insights into the long-runout of oceanographic gravity currents on mild slopes.

\end{abstract}


\section{Introduction}
\label{sec:Intro}
Inclined gravity currents are a type of wall-bounded buoyancy-driven shear flow \citep{simpson1999book}, serving as a critical yet poorly understood mechanism for the transport of various substances in geophysical and engineering environments. 
\cite{turner1959} were the first to study the dynamics of inclined gravity currents, using laboratory experiments in a sloping laboratory channel to show that the along-slope component of buoyancy in a current is resisted by drag owing to a combination of wall friction and entrainment of ambient fluid. This dynamic equilibrium determines the bulk flow speed in the current. 

\begin{figure}
  \centerline{\includegraphics[scale =0.5]{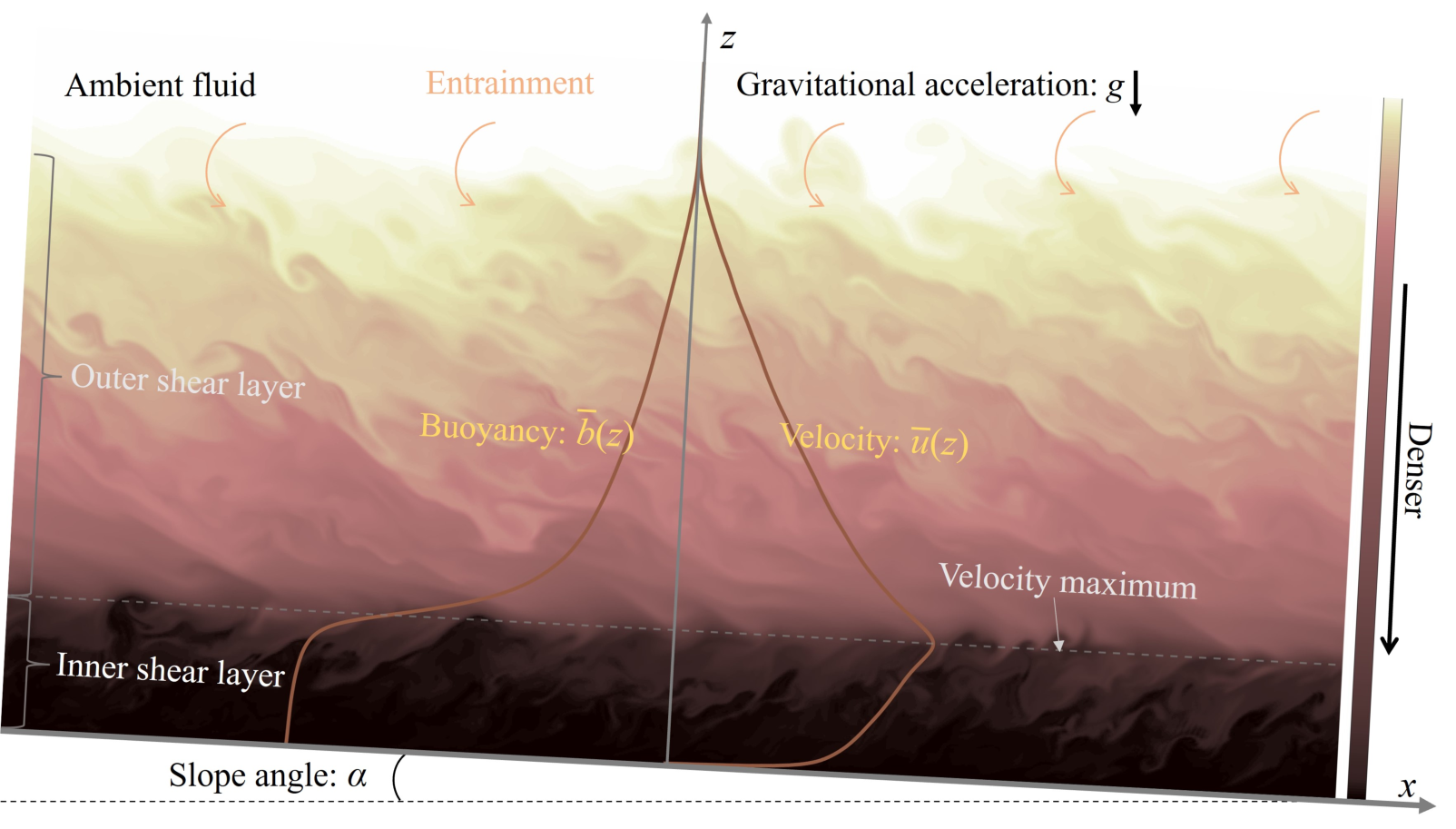}}
  \caption{Structure and instantaneous buoyancy field $b$ of an inclined gravity current. }
  \label{fig:system}
\end{figure}

Establishing a detailed understanding of the dynamics governing an inclined gravity current has proven challenging. In particular, the internal structure of a current generally consists of a relatively dense inner shear layer above the bottom boundary \citep[typically approximated by a boundary layer,][]{kneller1999innerWallslowdiffusionGaussian}, and an outer shear layer into which overlying ambient fluid is entrained at sufficiently large Reynolds number  \citep{turner1986assumption}. Figure \ref{fig:system} shows an example (from this study) of the instantaneous buoyancy field and internal structure in the body of an inclined gravity current. 
The two layers are naturally delineated by the level where the along-slope velocity reaches a maximum and where shear production of turbulent kinetic energy (TKE) must vanish. It is apparent, however, that different flow dynamics must govern the outer (free-shear-like) layer and the inner (boundary-layer-like) layer, resulting in differing growth rates and characteristic length scales. Moreover, the flows in each layer are coupled across the level of the velocity maximum.

The modelling of inclined gravity currents in a weakly stratified environment has been relatively well developed. These currents are characterized by a velocity maximum in close proximity to the wall, similar to a turbulent wall jet \citep{wei2021layeredjet}. Given the minimal role of the inner layer in these scenarios \citep{sequeiros2010gauprofile, luchi2018drivinglayer}, a scaling law based on the integral top-hat variables \citep{turner1959} of the overall current has been widely employed, analogous to the `outer scaling law' for a wall jet \citep{wygnanski1992ontheapp}. The flow variables normalized by the integral scales show considerable self-similarity at relatively large slope angles \citep{krug2013experimental, krug2015tur/nontur, krug2017fractal, van2018smallscale, van2019, dieu2020emily}. 

It is unclear if the integral top-hat formulation and disregard of the inner layer remain a valid approach for relatively strongly stratified currents on shallow-angled slopes. At decreasing angles, we expect an increasing portion of the current depth to be occupied by the inner layer as the driving component of the buoyancy forcing reduces. Indeed, there is accumulating evidence suggesting that the inner and outer layers become decoupled at small angles, driven by a range of underlying mechanisms. Examples include references to a `zone of strongly limited vertical turbulence' \citep{luchi2018drivinglayer}, `anti-diffusive mixing' \citep{selfsharp2019} and an `intermediate destruction layer' \citep{salinas2021nature}, all of which contribute to the formation of a transport barrier between the two layers.

In the present study, we conduct DNS of inclined gravity currents with no-slip bottom boundary conditions for a range of slope inclinations and initial Richarsdon numbers. Our aim is to investigate the internal structure and coupled dynamics that govern the behaviour of the currents. The outer layer in our simulations is compared with an inclined gravity current on a free-slip boundary \citep{van2019} because the boundary conditions are almost identical in both flows (apart from relaxation of the zero normal flux condition at the base of the outer layer). The inner layer in our simulations is compared with that in a turbulent planar channel flow, including both a closed channel \citep{CCFlee2015} and open channel \citep{OCFYao2022}. The ultimate objective of this paper is to develop a complete description of an inclined gravity current by matching the inner and outer layer solutions across the velocity maximum.

The set-up of the simulations and governing equations along with layer-specific parameters are outlined in \S\ref{sec:simulation}. In \S\ref{sec:temEvo}, we examine the evolution of the currents. A scaling model for the outer layer  is presented in \S\ref{sec:outer}. We then investigate the interactions between the outer and inner layers in \S\ref{sec:interaction} and develop a scaling model for the inner layer in \S\ref{sec:inner}. The inner-outer scaling models are matched in \S\ref{sec:matchCond} to describe the entire current and to model entrainment. Finally, we draw conclusions in \S\ref{sec:conclusion}.

\section{Case description}\label{sec:simulation}
\subsection{Simulation setup}
We consider a negatively buoyant gravity current flowing down a slope of constant angle $\alpha$, as shown in figure \ref{fig:system}. Periodic boundary conditions are imposed for all flow variables on the lateral boundaries of a finite-sized computational domain. Consequently, the simulations are statistically homogeneous in the streamwise $(x)$ and spanwise $(y)$ directions but evolve with time. The simulation setup follows the framework established by \cite{van2019}, with the exception of the bottom boundary condition, which in this study is specified as no-slip rather than free-slip. This setup leads to the evolution of a temporal gravity current, resulting in significant computational savings compared to simulations of a spatially evolving gravity current, especially for shallow angle cases involving a long evolution process. A detailed description of temporal gravity currents is provided in \cite{van2019}. 

If the flow is assumed Boussinesq, the governing equations in the coordinate system in figure \ref{fig:system} may be written as 
\begin{align}
  \label{eq:momentum equation}
  \frac{\partial \bf{u}}{\partial t}+{\bf u} \cdot \nabla {\bf u}&=-\rho_a^{-1}\nabla p+v\nabla^2{\bf u}
  +b\bf {e}, \\
    \label{eq:buoyancy transport}
  \frac{\partial b}{\partial t}+{\bf u} \cdot \nabla b &=\kappa\nabla^2 b, \\
  \label{eq:continuity equation}
  \nabla\cdot{\bf u}  &=0,
\end{align}
where $\bf{u}$ $ =(u,v,w)$ is the velocity vector, $p$ is the pressure, $b=(\rho_a-\rho)g/\rho_a$ is the buoyancy and $\rho_a$ is a reference density (taken to be that of the ambient fluid). The vertical unit vector resolved in the coordinate system is ${\bf e} = (-\sin{\alpha}, 0, \cos{\alpha})$, and $\nu$ and $\kappa$ are the kinematic viscosity and diffusivity, respectively. 

We solve numerically the governing equations using an in-house DNS code SPARKLE, which employs a conservative fourth-order-accurate differencing scheme \citep{verstappen2003numerics} for spatial discretization and an adaptive third-order Adams-Bashforth scheme for explicit time advancement. The code is described in detail by \cite{craske2015jet1} and has been widely used in simulations of gravity currents \citep{krug2017fractal,van2019, dieu2020emily}. The grid size $\Delta x$ of the domain varies between cases to ensure $\Delta x/\eta_K < 3/2$. Here, $\eta_K$ is a characteristic Kolmogorov length scale defined as $(\nu^3/\varepsilon_T)^{1/4}$, where $\varepsilon_T$ is a characteristic dissipation rate. Details of the simulations are listed in table \ref{tab:sim}.

\begin{table}
  \begin{center}
\def~{\hphantom{0}}
  \begin{tabular}{lccccccc}
      $Sim.$  & $\alpha$   & $Ri_0$ &$Ri_\infty$& $Re_\tau$&Resolution($x\cdot y\cdot z$)&$t_{ave}/t^*$ \\[3pt]
       1N ~&~1~  & ~1.11~&~0.36~& 740 &~$1536^3$&20 \\
       2N  & 2     & 0.56  & 0.25& 620  & ~$1536^3$&13\\
       5N & 5 &  0.22&0.18&260&~$1536^2\times1152$&13\\
       10N & 10 & 0.11&0.14&170 &~$1536^2\times1152$&13\\
       45N & 45 &  0.02&0.07&100 &~$1536^2\times 1024$&7\\
  \end{tabular}
  \caption{Simulation details. $Ri_0 = -b_0h_0\cos\alpha/u_0^2$ is the initial Richardson number, where $b_0, h_0$ and $u_0$ are the initial buoyancy, velocity and layer thickness, respectively. Note that $u_0$ and $h_0$ each maintain the same value across all simulations. $Ri_\infty$ represents the stabilized value of Richardson number $Ri$ when the flow is fully developed, where $Ri$ is defined in (\ref{eq:hubRi} $d$). The Reynolds number $Re_\tau=u_\tau z_{um}/\nu$ characterizes the inner layer, where $u_\tau=\sqrt{\left. \nu({\partial \overline{u}}/{\partial z})\right|_{z=0}}$ is the friction velocity and $z_{um}$ is the vertical coordinate of the velocity maximum. $t_{ave}$ is a time interval towards the end of the simulation over which the numerical results are averaged and $t^* = h_0/\sqrt{B_0}$ is a typical timescale. The initial Reynolds number $Re_0 = u_0h_0/\nu$ is 3800 in all simulations. The size of the computational domain is $20h_0\times20h_0\times20h_0$ for all cases.}
  \label{tab:sim}
  \end{center}
\end{table}

\subsection{Characteristic quantities}

Given the statistical homogeneity in the $x$ and $y$ directions, we spatially average the governing equations and write the Reynolds-averaged momentum and buoyancy equations as
\begin{align}
 &\frac{\partial\overline{u}}{\partial t} +\frac{\partial\overline{w'u'}}{\partial z}=v\frac{\partial^2\overline{u}}{\partial z^2}-\overline{b}\sin\alpha, 
 \label{eq:averaged momentum} \\
 &\frac{\partial\overline{b}}{\partial t} + \frac{\partial\overline{w'b'}}{\partial z}=\kappa\frac{\partial^2\overline{b}}{\partial z^2},
  \label{eq:averaged buoyancy} 
\end{align}
where $\overline{*}=\iint *~\d x \d y/(L_xL_y)$ represents the spatial averaging operator for the quantity *, $L_x$ and $L_y$ are the dimensions of the domain in the $x$ and $y$ directions, respectively, and a prime  represents the departure from the corresponding average, i.e. $*' = * - \overline{*}$. Taking the dot product of $\bf{u}$ with the momentum equation \eqref{eq:momentum equation}, subtracting the mean kinetic energy ($\overline{\bf u}\cdot \overline{\bf u}/2$) and averaging over $x$ and $y$ directions, we obtain the TKE budget
\begin{equation}
     \frac{\partial e}{\partial t}  = - \overline{w'u'} \frac{\partial \overline u}{\partial z} + \overline{w'b'} \cos \alpha -  \overline{u'b'} \sin \alpha - \varepsilon,
  \label{eq:averaged TKE}
\end{equation}
where $e=\overline{{u'_i}^2}/2$ is the TKE, and $\varepsilon=\nu\overline{(\partial u_i'/\partial x_j)^2}$ is the dissipation rate of TKE. Note that the TKE transport terms are neglected.

The integral volume flux $Q$, momentum flux $M$ and integral buoyancy forcing $B$ of the gravity current are defined here as
\begin{equation*}
    Q=\int_0^\infty\overline{u}dz,\quad  M=\int_0^\infty\overline{u}^2dz,\quad B=\int_0^\infty -\overline{b}\sin{\alpha}dz.
    \refstepcounter{equation}
   \eqno{(\theequation{a-c})}
\end{equation*}
We decompose these quantities into inner and outer components denoted with subscripts $i$ and $o$, respectively, i.e.\ 
\begin{equation*}
\begin{aligned}
     Q =&\underbrace{\int_0^{z_{um}}\overline{u}dz}_{Q_i} + \underbrace{\int_{z_{um}}^\infty \overline{u}dz}_{Q_o},\quad M =  \underbrace{\int_0^{z_{um}}\overline{u}^2dz}_{M_i} + \underbrace{\int_{z_{um}}^\infty \overline{u}^2dz}_{M_o}\\
     B=& \underbrace{\int_0^{z_{um}}-\overline{b}\sin{\alpha}dz}_{B_i} + \underbrace{\int_{z_{um}}^\infty -\overline{b}\sin{\alpha}dz}_{B_o},
\end{aligned}
   \refstepcounter{equation}
   \eqno{(\theequation{a-c})}
    \label{eq:QMB} 
\end{equation*}
where $z_{um}$ is the vertical coordinate of the velocity maximum. Note that buoyancy is conserved in the flow and thus $B$ is constant (equal to $B_0$). We further keep $B$ $(=B_0)$ constant across the series of slope angles considered by adjusting the initial buoyancy field in each simulation. The characteristic velocity scale $u_{T*}$, layer thickness $h_*$ and buoyancy $b_{T*}$ and the bulk Richardson number $Ri_*$ are defined as 
\begin{equation*}\label{eq:hubRi}
    h_*=\frac{Q_*^2}{M_*},~ u_{T*}=\frac{Q_*}{h_*},~ b_{T*}=-\frac{B_*}{h_*\sin\alpha},~ Ri_*=\frac{-b_{T*}h_*\cos\alpha}{u_{T*}^2}=\frac{B_*}{u_{T*}^2\tan\alpha},
    \refstepcounter{equation}
    \eqno{(\theequation{a-d})}
\end{equation*}
respectively, where the subscript $*$ is either omitted, $i$ or $o$ and is used to characterize the entire current, the inner layer or the outer layer, respectively. In a similar manner, the characteristic scales for TKE $e_{T*}$ are given by
\begin{equation}
     \underbrace{\int_0^\infty e dz}_{e_Th} = \underbrace{\int_0^{z_{um}} e dz}_{e_{Ti}h_i} + \underbrace{\int_{z_{um}}^\infty e dz}_{e_{To}h_o}.
     \label{eq:e_T}
\end{equation}
\section{Temporal evolution} \label{sec:temEvo}
\subsection{Inner-outer flow dynamics} \label{sec:IOdynamics}
\begin{figure}
  \centering
   \includegraphics[scale =0.25, trim=0.3cm 0.2cm 0cm 0.4cm, clip] {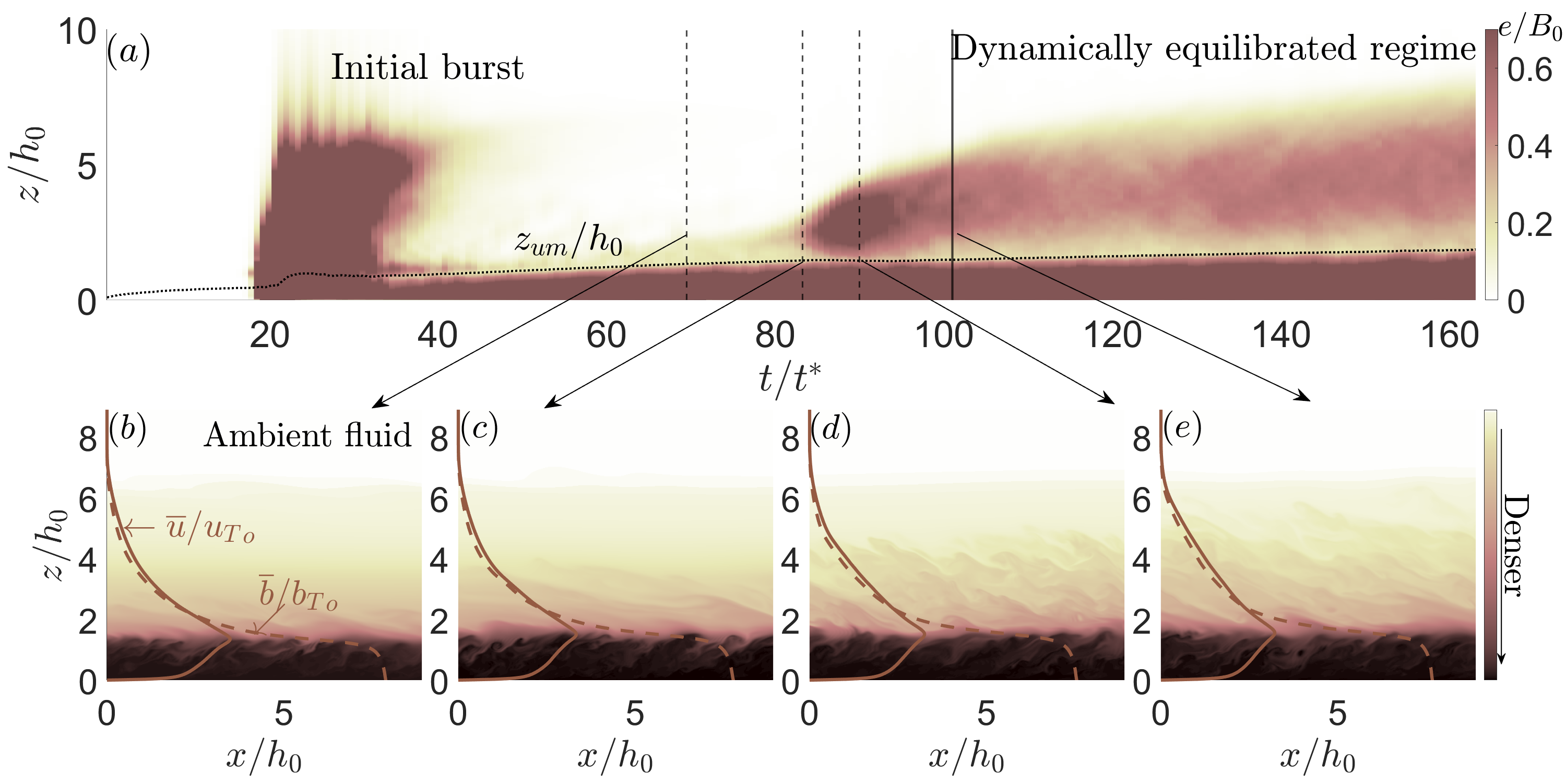}
\includegraphics[scale =0.25,trim=0cm 0.5cm 1.1cm 0cm, clip]{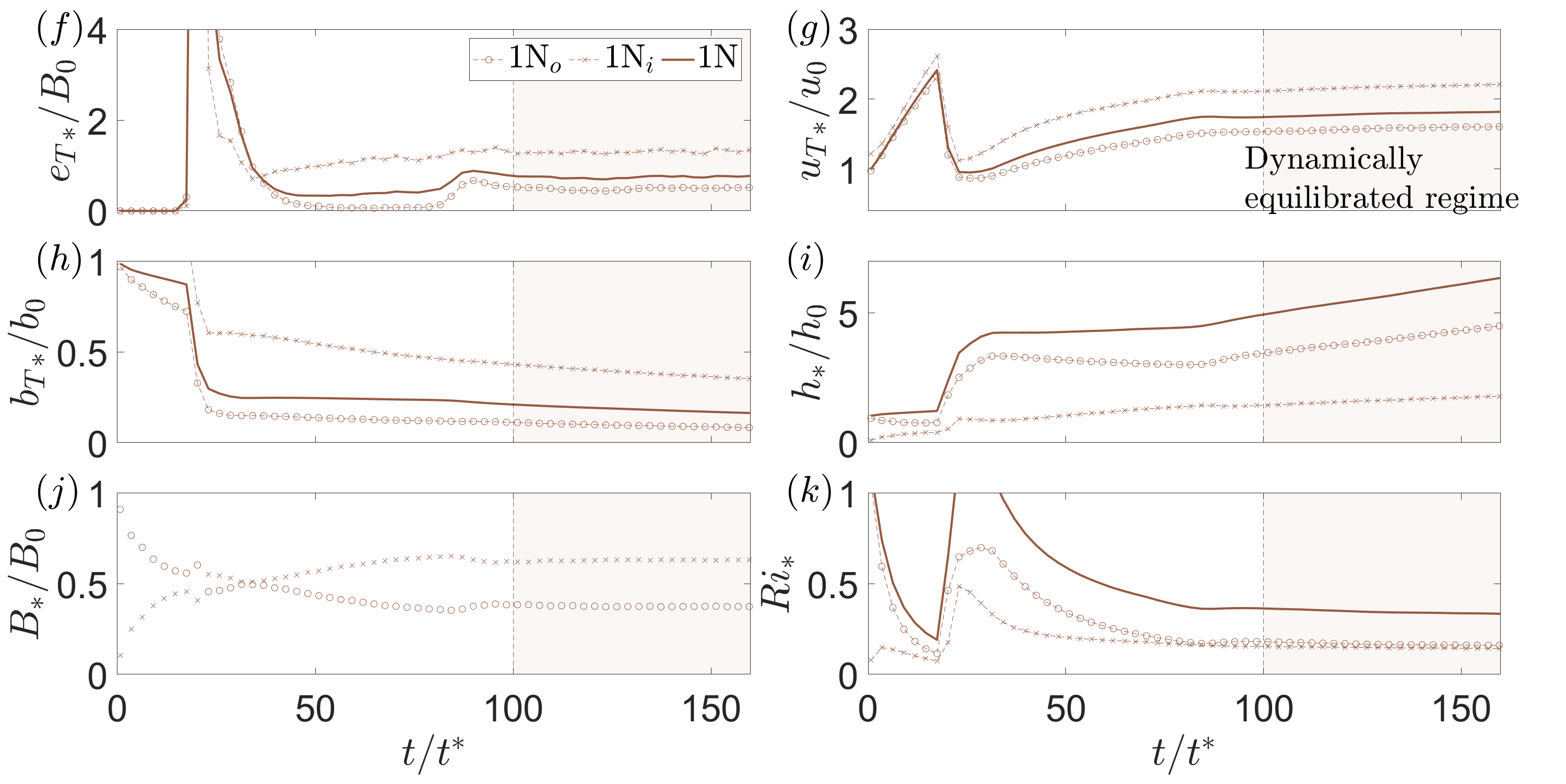}
  \caption{Temporal evolution of  $(a)$ dimensionless TKE: $e/B_0$ for 1N, where the dotted line denotes the boundary between the inner and the outer layers, together with the instantaneous buoyancy field and profiles of horizontally-averaged velocity and buoyancy, $\overline{u}/u_{To}$ and $\overline{b}/b_{To}$, at $(b) t/t^* = 69.5, (c) t/t^* = 83.25, (d) t/t^* = 90$ and $(e) t/t^* = 101$. Temporal evolution of normalized $(f) e_T, {e_T}_i,{e_T}_o$; $(g) u_T, {u_T}_i, {u_T}_o$; $(h) b_T, {b_T}_i, {b_T}_o$; $(i) h, h_i, h_o$; $(j)B_i, B_o$ and $(k) Ri, Ri_i, Ri_o$. Note that the subscript '$o$' and '$i$' in the legend denote the results of the outer layer (displayed with symbol '\scalebox{1.5}{$\circ$}') and the inner layer  (displayed with symbol '$\times$'), respectively.}
 \label{fig:evolution}
\end{figure}

As we observe an essentially similar evolution processes for all the slope angles considered, we use case 1N as an example to illustrate the dynamics.
Figure \ref{fig:evolution} $(a)$ shows the temporal evolution of $e(z,t)$, normalized by the integral buoyancy forcing $B_0$, against $t/t^*$ for case 1N, where $t^*=h/\sqrt{B_0}$ is a typical timescale. Also shown is the location of the velocity maximum $z_{um}$ normalized by $h_0$ (black dashed line). Instantaneous snapshots at different time intervals of the normalized buoyancy field  in figures \ref{fig:evolution} $(b-e)$ show the development of  turbulent structures in the flow. The evolution of layer-specific characteristic flow variables for case 1N are presented in figure \ref{fig:evolution} $(f-k)$; as above, an omitted subscript, $i$ or $o$ is used to denote the overall current, the inner layer and the outer layer, respectively. 

An intense initial burst of turbulence associated with shear instabilities is observed for $20<t/t^*<40$ in figure \ref{fig:evolution} $(a)$,  caused by the sharp initial acceleration (figure \ref{fig:evolution} $(g)$) from the initial conditions. This initial burst leads to a noticeable plunge in velocity (see figure \ref{fig:evolution} $(g)$), as the mean flow kinetic energy is converted to TKE and potential energy (see the increase of $h_*$ in figure \ref{fig:evolution} $(i)$). Consequently, large bulk Richardson numbers arise after the initial burst (see figure \ref{fig:evolution} $(k)$) as damping of turbulence and even relaminarization occurs in the outer layer (see Figure \ref{fig:evolution} $(a)$, $50<t/t^*<80$). Meanwhile, turbulence is sustained in the inner layer over the whole evolution process. 

During the period of damping ($50<t/t^*<80$), the outer layer again accelerates (figure \ref{fig:evolution} $(g)$) due to the buoyancy forcing, leading to increased shear and reduced Richardson number (see figure \ref{fig:evolution} $(k)$). The outer layer eventually transitions to a turbulent state as the shear instabilities overcome the restoring stratification. The temporal sequence of instantaneous buoyancy fields (figure \ref{fig:evolution} $(b)$ -- $(e)$) illustrates the transition to a turbulent regime in the outer layer, which initiates with the onset of instabilities near the velocity maximum, followed by the growth of eddies and vortices. Figure \ref{fig:evolution} $(f)$ quantifies the turbulence level throughout the evolution process, showing the first burst of turbulence and subsequent damping (confined to the outer layer), followed by the eventual transition to  a nearly constant turbulence level. 

Restricting our attention to the time period  $t/t^*>100$, it is noteworthy that the characteristic velocities ${u_T}_i$, ${u_T}_o$, and $u_T$ attain nearly constant values (see figure \ref{fig:evolution} $(f)$). This behaviour is consistent with the so-called equilibrium state commonly assumed to exist for inclined gravity currents, in which a bulk force balance is achieved \citep{turner1959, britter1980motion, odier2014linearity, martin_ex_2019}.

The buoyancy variables ${b_T}_i$, ${b_T}_o$ and $b_T$ shown in figure \ref{fig:evolution} $(h)$ continue to reduce gradually as entrainment of ambient fluid continues to dilute the current and increase the layer thickness (figure \ref{fig:evolution} $(i)$). However, the integral buoyancy forcings $B_o$ and $B_i$ (figure\ref{fig:evolution} $(j)$) are remarkably invariant, i.e.\ the total buoyancy in each of the inner and the outer layers is approximately conserved. This behaviour is closely linked to the interaction between the two layers, and is discussed in \S\ref{sec:ExOfMaB}. 

The bulk Richardson numbers shown in figure \ref{fig:evolution} $(k)$ also attain approximately constant values in the turbulent regime. We will, therefore, refer to this regime as dynamically equilibrated,  and the flow behaviour in the outer and the inner layers are described in detail in \S\ref{sec:outer} and \S\ref{sec:inner}, respectively. The flow quantities in the outer layer exhibit strong self-similarity during the dynamically equilibrated regime, and hence we will refer to this regime in the outer layer as the self-similar regime.

\subsection{Slope angle dependence} \label{sec:angleDependence}

\begin{figure}
  \centering
\includegraphics[scale =0.25,trim=0cm 0.5cm 1.1cm 0.4cm, clip]{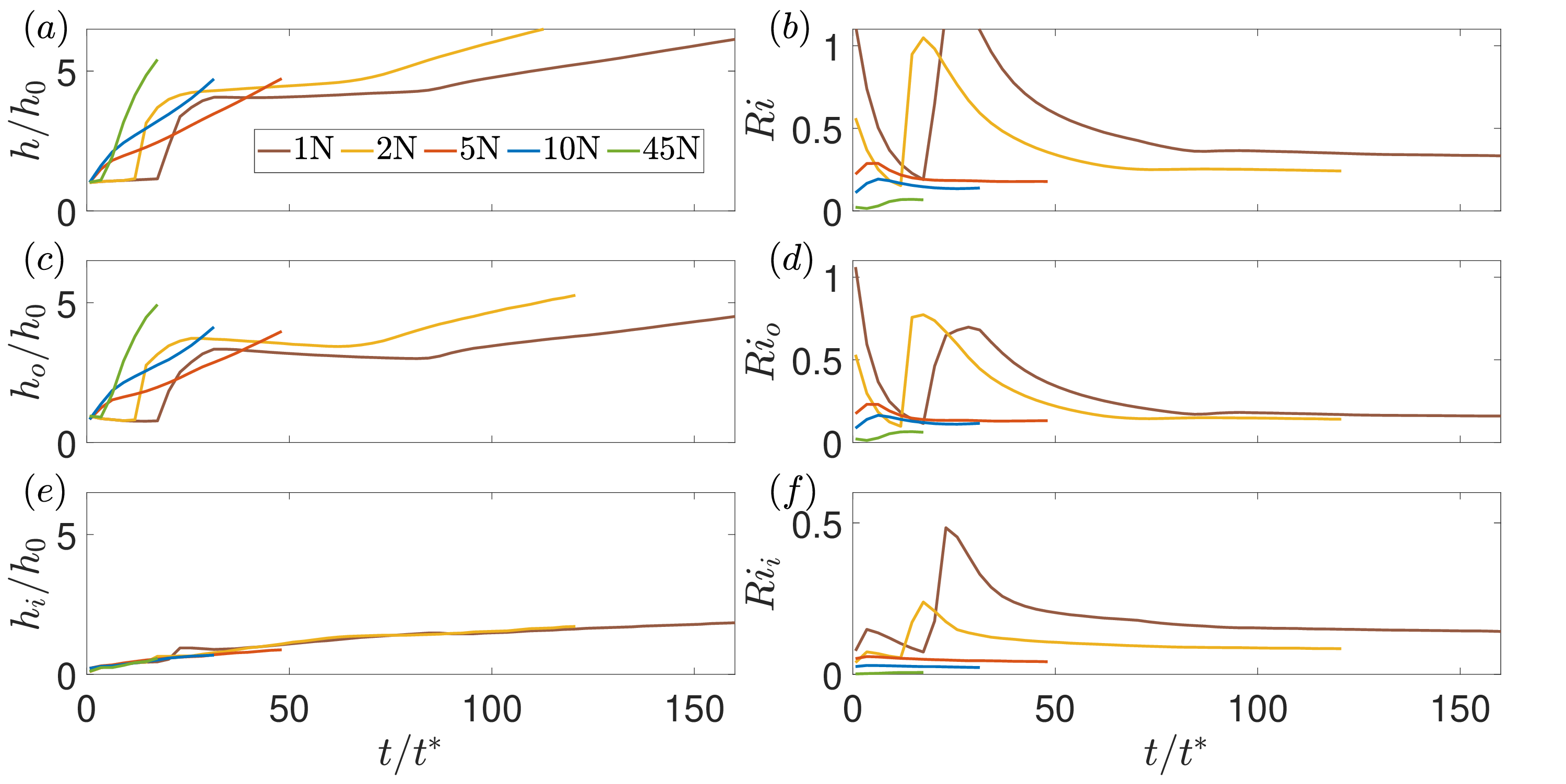}
  \caption{Temporal evolution of overall quantities: $(a) h/h_0 $ and $(b) Ri$; outer layer quantities: $(c) h_o/h_0 $ and  $(d)Ri_o$; inner layer quantities: $(e) h_i/h_0$ and $(f) Ri_i$.}
  \label{fig:evolutionOfAll}
\end{figure}

Figures \ref{fig:evolutionOfAll} $(a), (c)$ and $(e)$ show the evolution of the characteristic thicknesses ($h, h_o, h_i$) of the currents for different slope angles. The overall thickness $h$ (see figure \ref{fig:evolutionOfAll} $a$) and outer layer thickness $h_o$ (see figure \ref{fig:evolutionOfAll} $c$) grows more rapidly on steeper slopes because of relatively vigorous entrainment. Notably, the overall current and the outer layer exhibit a similar normalized growth rate (approximately linear in time), whilst the inner layer exhibits much slower growth rate over time compared with the outer layer. 

Figures \ref{fig:evolutionOfAll} $(b), (d)$ and $(f)$ plot the bulk Richardson numbers ($Ri, Ri_o, Ri_i$) as a function of time for different slope angles. In all cases, the various Richardson numbers become essentially constant in the self-similar regime and are negatively correlated with slope angle, suggesting a stronger stratification at a shallower angle. 

\begin{figure}
  \centerline{\includegraphics[scale =0.33, trim=0.8cm 12.5cm 0.8cm 1.1cm, clip]{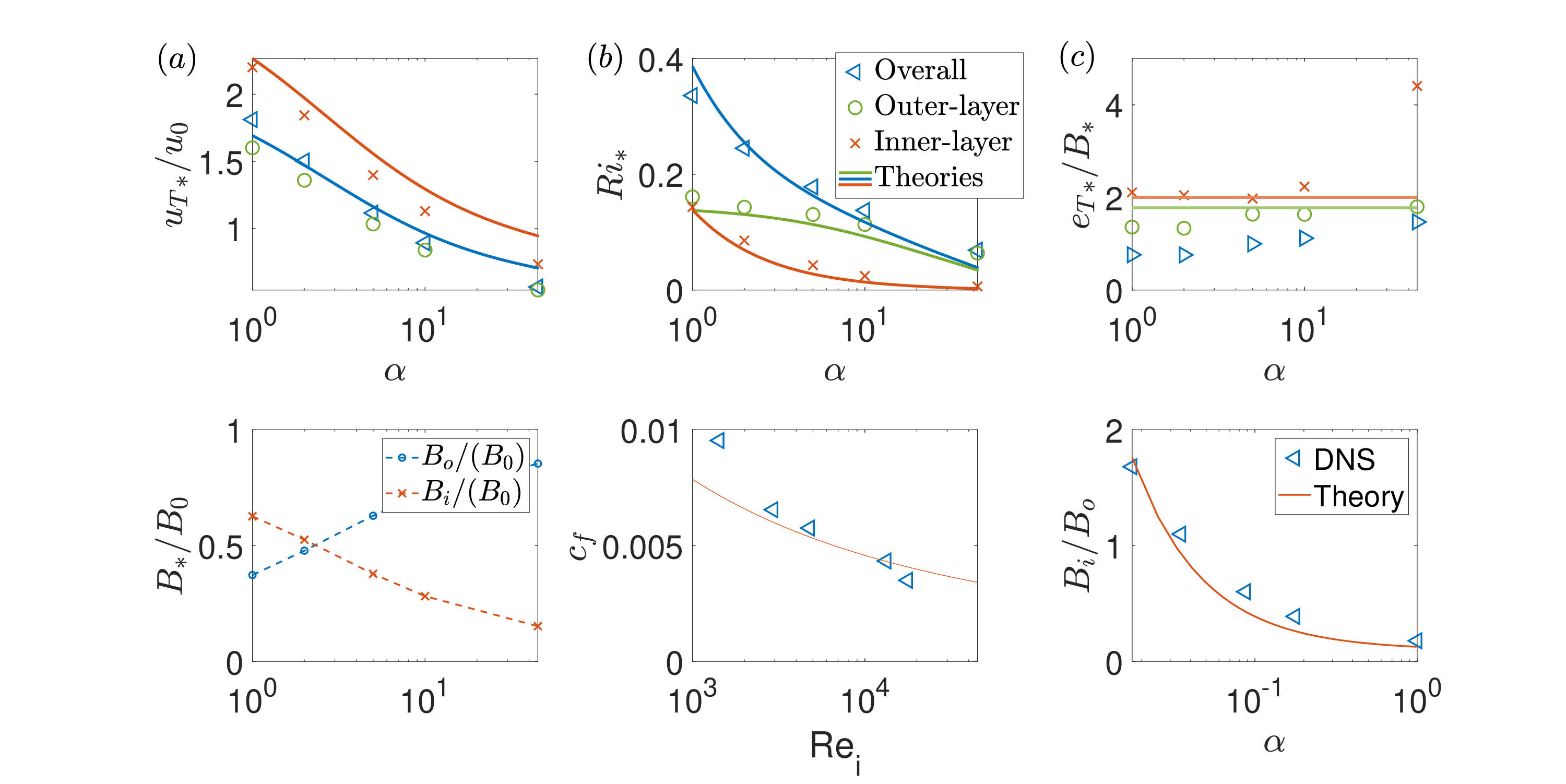}}
  \caption{Variation of averaged $(a)$ normalized $u_T$, $(b) Ri,$ and $(c)$ normalized $e_T$ over $t_{ave}$ in the dynamically equilibrated regime against $\alpha$ in degrees, including the overall, outer and inner quantities for all the slope angles considered. The solid lines in panels $(a)$ and $(b)$ denote the theoretical predictions from \eqref{eq:umSolu} and \eqref{eq:Risolu}, respectively. The solid lines in panel $(c)$ represent the prediction in (\ref{eq:outereoB}) and (\ref{eq:innereToBi}).}
\label{fig:averagedVars}
\end{figure}

Figure \ref{fig:averagedVars} shows the overall and layer-averaged  normalized   velocities, Richardson numbers and normalized TKE  (the subscript $*$ denoting whether the quantity is overall or layer-specific) for the currents as a function of slope angle in the self-similar regime. The characteristic velocities $u_{T*}$ shown in figure \ref{fig:averagedVars} $(a)$ attain larger values at shallower angles ($u_0$ is set to the same value across all the slope angles), reflecting the need for greater shear to overcome stronger stratification and transition the current to the turbulent state. We also observe that $u_T\approx u_{To}$ for all the slope angles considered.


Figure \ref{fig:averagedVars} $(b)$ shows that the Richardson numbers in the dynamically-equilibrated state increase as the slope angle decreases. $Ri_o$ appears to approach an asymptotic value at a small angle (note the logarithmic scale). This is consistent with the conjecture that stratified shear flows adjust to a `marginally stable' state \citep{turner1979buoyancyeffects} characterized by a critical Richardson number. The overall and inner Richardson numbers, $Ri$ and $Ri_i$, respectively, increase rapidly as the  slope angle decreases (see \S\ref{sec:innerProfiles} and further discussion in \S\ref{sec:matchCond}). The   normalized TKE is seen in figure \ref{fig:averagedVars} $(c)$ to be approximately constant at the three smallest angles (1, 2 and 5 degrees) in each of the outer and inner layers. 


\section{Outer-shear-layer scaling}\label{sec:outer}
\subsection{Self-similar profiles}\label{sec:outerProfiles}
\begin{figure}
  \centerline{\includegraphics[scale =0.29, trim=0.2cm 0cm 1cm 0.9cm, clip]{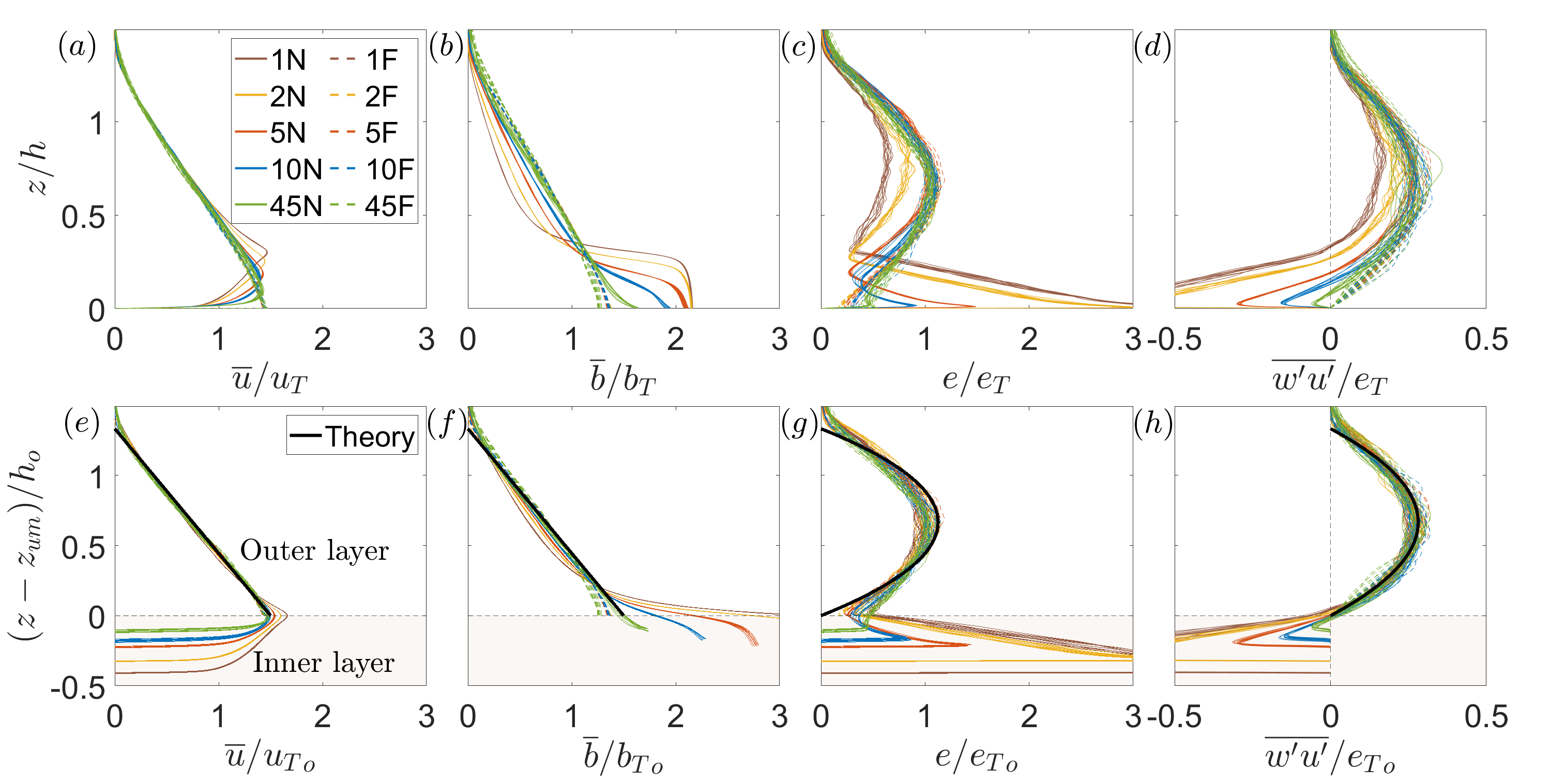}}
 \caption{Profiles of $(a)$ $\overline{u}/u_T$, $(b)$ $\overline{b}/b_T$, $(c) e/e_T$ and $(d) \overline{w'u'}/e_T$ against scaled height $z/h$; $(e)$ $\overline{u}/{u_T}_o$, $(f)$ $\overline{b}/{b_T}_o$, $(g) e/{e_T}_o$ and $(h) \overline{w'u'}/e_{To} $ against scaled distance from the velocity maximum $(z-z_{um})/h_o$ (outer layer scaling). Profiles for each case at a series of times in the self-similar regime are plotted. Strong self-similarity and collapse of profiles are observed in the outer layer for all the cases considered when normalization is based on  outer layer integral quantities. The results for no-slip boundaries (from this study), and free-slip boundaries(adapted from \citep{van2019}) are appended with "N" and "F," respectively in the legend (e.g., 5F for a 5-degree slope with free-slip boundaries). The sold lines in panels $(e-h)$ represent the predictions from \eqref{eq:fufbfe} and \eqref{eq:outer_wucme}.} 
\label{fig:outerProfiles}
\end{figure}

Figure \ref{fig:outerProfiles} shows  normalized profiles of velocity $\overline{u}$, buoyancy  $\overline{b}$, TKE $e$ and turbulent shear stress $\overline{w'u'}$ during the dynamically equilibrated regime for all the cases considered, including both no-slip (solid lines) and free-slip (dashed lines) boundary conditions. The results for the free-slip boundary conditions are from \cite{van2019}. Each profile is scaled by the appropriate integral quantity ($u_T$, $b_T$ or $e_T$) at the time of sampling in the self-similar regime to give  figure \ref{fig:outerProfiles} $(a-d)$. Note that $\overline{w'u'}$ is scaled with $e_T$, and this is discussed in \S\ref{sec:paraTKE}. The normalized profiles for the flow variables largely collapse for all free-slip cases, whereas deviations become evident  with no-slip boundary conditions. 

The observed deviations suggest use of a local scaling based on the  integral flow quantities in the outer layer (i.e.\ ${u_T}_o$, ${b_T}_o$ and ${e_T}_o$).
Figure \ref{fig:outerProfiles} $(e-g)$ shows the profiles of velocity, buoyancy, TKE and Reynolds stress rescaled with the appropriate outer layer integral quantities. In addition, a normalized vertical (`outer') coordinate $(z-z_{um})/h_o$ is used to facilitate meaningful comparison between the results for free-slip and no-slip boundary conditions (i.e.\ the vertical coordinate has its origin at the level of the maximum velocity for both types of boundary conditions). 
Remarkably, all the normalized profiles  nearly collapse in the outer layer, indicating that similar dynamics dominate there, whether or not an inner layer (corresponding to the region $(z-z_{um})/h_o < 0$) is present. Although there are some deviations from universal forms, these look to be 
associated primarily with the presence of a strong stratification localized near the velocity maximum for currents on a  low angled no-slip boundary  (see \S\ref{sec:innerProfiles} for more details).

\subsection{Approximate self-similar solutions} \label{sec:outerSolutions}
 Given the observed near-collapse of the outer-layer profiles in \S\ref{sec:outerProfiles}, 
 we now examine the usefulness of the approximate self-similar descriptions developed by \cite{van2019} for inclined currents on a free-slip boundary.
 We thus propose that the outer layer profiles are modelled as
\begin{equation*}
\overline{u}=\underbrace{a_uB_o^{1/2}}_{u_{To}} \frac{2}{\eta_1^2}(\eta_1 - \eta_o), \quad
\overline{b}=\underbrace{-a_b\frac{B_o}{h_o\sin\alpha }}_{b_{To}} \frac{2}{\eta_1^2}(\eta_1 - \eta_o), \quad e = \underbrace{a_eB_o}_{e_{To}}\frac{6 \eta_o}{\eta_1^3}(\eta_1 - \eta_o),
\label{eq:fufbfe}
\refstepcounter{equation}
  \eqno{(\theequation{a-c})}
\end{equation*}
where the outer-layer self-similarity variable is defined as $\eta_o = (z-z_{um})/h_o \in[0,\eta_1]$ 
and the shape factor $\eta_1 = 4/3$. Note that equation (\ref{eq:fufbfe} $a-c$) reduces to the form considered by \cite{van2019} for a current on a free-slip boundary upon setting $z_{um} = 0$ and dropping the subscript $o$.

The coefficients $a_u$, $a_b$ and $a_e$ depend on the dimensionless parameters of the problem and need to be determined.  We adopt the results of \cite{van2019}, who observed that the eddy viscosity, eddy diffusivity, dissipation rate and turbulent Prandtl number could be parameterized as
\begin{equation*}
K_m = c_m \frac{e}{S}, \quad\quad 
K_\rho = c_\rho \frac{e}{S}, \quad\quad 
\varepsilon = c_\varepsilon e S,
\quad{\rm and}\quad Pr_T=\frac{c_m}{c_\rho},
\label{eq:shearScaling}
\refstepcounter{equation}
\eqno{(\theequation{a-d})}
\end{equation*}
respectively, where $S=|\partial\overline{u}/\partial z|$ is the absolute strain rate of the mean flow and  $c_m=0.25$, $c_\rho=0.31$ and $c_\varepsilon=0.21$ are empirical coefficients based on the DNS results. It follows from the success of these scalings in terms of the strain rate $S$ and TKE $e$ that the turbulence is in the shear-dominated regime \citep{mater2014threeregime, krug2017fractal}. 
Armed with this turbulence closure, \cite{van2019} integrated the equations for Reynolds-averaged momentum, buoyancy 
and turbulent kinetic energy, \eqref{eq:averaged momentum}-\eqref{eq:averaged TKE}, and used the Von Pohlhausen method \citep{lighthill1950, spalding1954, schlichting2016} to find the coefficients 
\begin{equation*}
 a_u = \left(\frac{9}{8}\frac{Pr_T(c_m-c_\varepsilon)\tan \alpha}{\tan\alpha Pr_T + c_m} \right)^{-1/2}, \quad\quad a_b=1.
 \label{eq:mvrSolu}
 \refstepcounter{equation} 
 \eqno{(\theequation{a,b})}
\end{equation*}
Note, however, that the coefficient for the turbulent kinetic energy $a_e$ (see equation \ref{eq:fufbfe} $c$) did not follow from this analysis and we evaluate it in the next section.

The theoretical solutions given by equations (\ref{eq:fufbfe}--\ref{eq:mvrSolu}) are shown in figure \ref{fig:outerProfiles} $(e-h)$, and are in good agreement with data from the simulations conducted in this study. Notably, the theory predicts that $e$ will tend to zero near $\eta_o = 0$  (i.e.\ $z=z_{um}$), but because this is not a solid boundary in these simulations, the TKE does not have to be zero there. Despite this, the shear production of TKE is zero at the velocity maximum by definition and the magnitude of $e/e_{To}$ is indeed close to zero. Therefore, the theory still provides a reasonable approximation.

\begin{figure}
  \centerline{\includegraphics[scale =0.29, trim=0cm 12.5cm 2.5cm 1.4cm, clip]{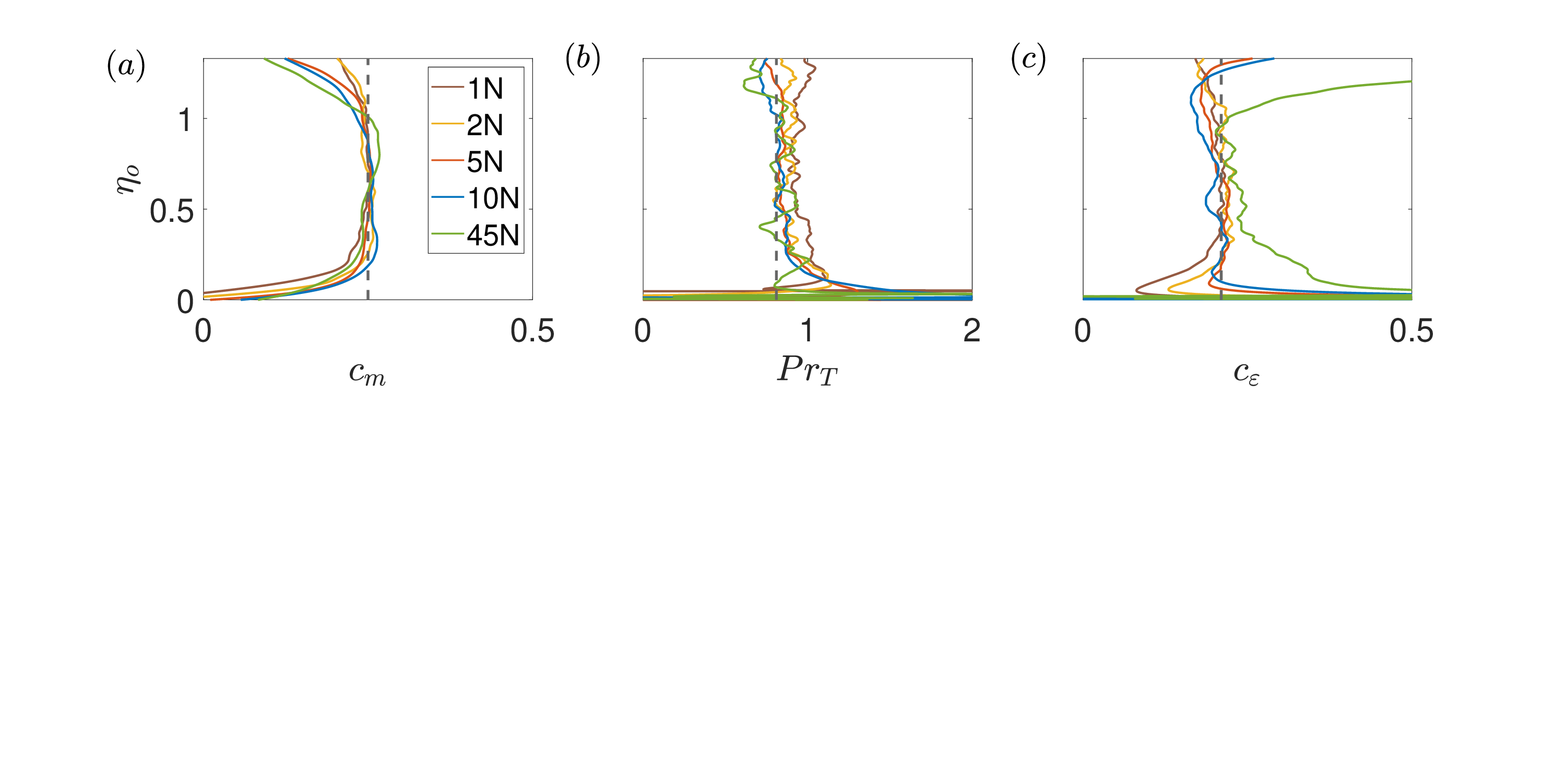}}
  \caption{Outer layer scaling of averaged turbulence parameters over $t_{ave}$ in the self-similar regime: $(a) c_m=K_mS/e\approx0.25, (b) Pr_T=c_m/c_{\rho}\approx0.81$ (i.e.\ $c_{\rho} \approx 0.31$) and $(c) c_\varepsilon=\varepsilon/(eS)\approx0.21$ against scaled distance to velocity maximum $(z-z_{um})/h_o$. The converged values are denoted with the vertical dashed lines.} 
\label{fig:TurParOuter}
\end{figure}

Figures \ref{fig:TurParOuter} $(a-c)$ show the temporally averaged scaling coefficients defined in \eqref{eq:shearScaling} over times sampled during the self-similar phase. We observe that there is a convincing collapse of profiles across the range of angles, with the converged values matching those from the free-slip cases. We thus conclude that the theory developed by \cite{van2019} can be effectively applied to the outer layer of an inclined current on a no-slip boundary.

\subsection{Scaling of turbulent kinetic energy } \label{sec:paraTKE} 
In this section, we explore if the TKE can be scaled with the integral buoyancy forcing $B$ as suggested by \cite{van2019}, thereby allowing all the turbulence quantities in the closure to be related to macroscopic flow quantities. We first assume that the turbulent shear stress $\overline{w'u'}$ can be parameterized in the outer layer (where $\partial \overline u / \partial z < 0$) using the gradient diffusion hypothesis and equation \eqref{eq:shearScaling}:
\begin{equation}
\overline{w'u'}= -K_m \frac{\partial \overline u}{\partial z} = c_m \frac{e}{S} S = c_m e.
\label{eq:wucme_outer}
\end{equation}
Substituting for $e$ using equation (\ref{eq:fufbfe} $c$), we expect $\overline{w'u'}$ to take a self-similar form
\begin{equation}
\overline{w'u'}=c_m{e_T}_o\frac{6\eta_o}{\eta_1^3}\left(\eta_1-\eta_0\right),
    \label{eq:outer_wucme}
\end{equation}
which is plotted in figure \ref{fig:outerProfiles} $(h)$ and shows good agreement with the DNS data. Note that, in contrast with $e$, we observe that $\overline{w'u'}$ does become zero at $\eta_o=0$. This is because $z_{um}$ is defined via the velocity maximum where $\partial \overline{u}/\partial z = 0$, and the gradient diffusion hypothesis (see \ref{eq:wucme_outer}) works reasonably well. Thus, the quadratic profile (\ref{eq:outer_wucme}) is more appropriate for $\overline{w'u'}$ than for $e$. 

\begin{figure}
  \centerline{\includegraphics[scale =0.29, trim=2.5cm 12.5cm 2.5cm 0.2cm, clip]{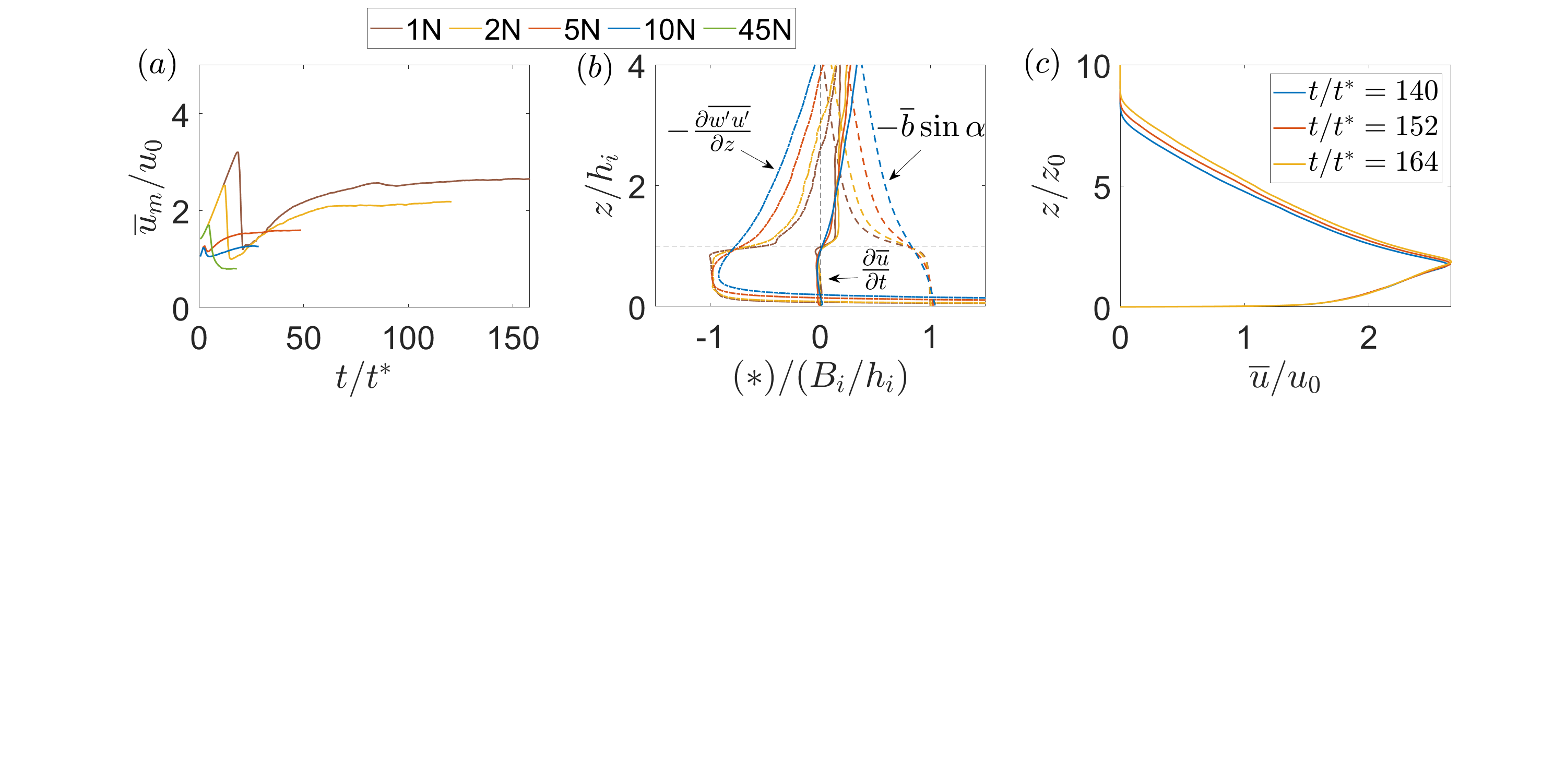}}
   \caption{$(a)$ Temporal evolution of the maximum velocity $\overline{u}_m$; $(b)$ momentum budgets averaged in the dynamically equilibrated regime over $t_{ave}$ for cases 1N, 2N, 5N and 10N; $(c)$ the instantaneous along-slope velocity profiles in the dynamically equilibrated regime at $t/t^*=140, 152, 164$ for case 1N. The top-left legend applies to panels $(a)$ and $(b)$.}
 \label{fig:dynamics}
\end{figure}

In order to find $a_e (= e_{To}/B_o)$, we first substitute the self-similar expressions (\ref{eq:fufbfe} $a$) and \eqref{eq:outer_wucme} into the streamwise momentum equation (\ref{eq:averaged momentum}) to give
\begin{equation}
  \frac{\partial \overline{u}}{\partial t}=\underbrace{\frac{12c_m{e_T}_o-2B_o\eta_1}{{h_o}\eta_1^3}\eta_o + \frac{2B_o\eta_1-6c_m{e_T}_o}{(h_o)\eta_1^2}}_{-\partial \overline{w'u'}/\partial z-\overline{b}\sin\alpha}.
    \label{eq:MomPar}
\end{equation}
Secondly, a crucial simplification is motivated by the observation that the characteristic velocity in the outer-layer $u_{To}$ becomes approximately constant (or only evolves over a relatively large time scale, see figure \ref{fig:evolution}) in the self-similar regime, consistent with the theory described in \cite{van2019}. Therefore, the maximum velocity in the outer layer $\overline{u}_m=3u_{To}/2$ (see \ref{eq:fufbfe} $a$) is also expected to become approximately constant in the self-similar regime, consistent with Figure \ref{fig:dynamics} $(a)$. 

Setting $\partial \overline{u}/\partial t \approx 0$ at $\eta_o=0$ in  (\ref{eq:MomPar}) gives $2B_o\eta_1-6c_m{e_T}_o=0$, and thus
\begin{equation}
a_e \equiv\frac{{e_T}_o}{B_o}= \frac{\eta_1}{3c_m} \approx 1.77,
    \label{eq:outereoB}
\end{equation}
and
\begin{equation}
    \frac{\partial \overline{u}}{\partial t}=\frac{9B_o}{8h_o}\eta_o,\quad \eta_o \in [0, \eta_1].
    \label{eq:puptfinal}
\end{equation}
The prediction from \eqref{eq:outereoB} is shown in figure \ref{fig:averagedVars} ($c$). Although the shear-dominated scaling mainly applies in the core region ($\eta_o\in[0.5, 1]$) of the outer layer, there is fairly good agreement with the DNS data over the entire outer layer. 

Equation \eqref{eq:puptfinal} predicts that the acceleration ${\partial\overline{u}}/{\partial t}$ increases with height above the velocity maximum, and is consistent with the instantaneous velocity profiles in figure \ref{fig:dynamics} $(c)$. Analysis of the momentum budget shows that the individual terms in \eqref{eq:MomPar} also vary linearly with height above the velocity maximum in the outer layer (figure \ref{fig:dynamics} $(b)$). Notably, the gradient of the Reynolds stress and the buoyancy terms are in approximate balance (i.e.\ no acceleration) at the velocity maximum and in the inner layer. This observation is discussed in detail in \S\ref{sec:inner}.

\section{Inner-outer-layer interaction}\label{sec:interaction}
In this section, we demonstrate that the inner and outer layers  are weakly coupled, providing a theoretical foundation for the development of layer-specific scaling laws.

\subsection{Integral momentum and buoyancy budgets} \label{sec:ExOfMaB}

The integral momentum and buoyancy equations for the inner and outer layers can be obtained by integrating \eqref{eq:averaged momentum} and \eqref{eq:averaged buoyancy} over the respective layer to give
\begin{equation*}  
\begin{aligned}
      \frac{\d B_o}{\d t} &= f_{um} \sin{\alpha},\qquad\qquad\quad &
\frac{\d B_i}{\d t} &= -f_{um} \sin{\alpha}, \\
        \frac{\d Q_o}{\d t} &= B_o + m_{um}, &
\frac{\d Q_i}{\d t} &= B_i -\tau_w -m_{um}, 
\end{aligned}
\label{eq:IntMomBuoBud} 
  \refstepcounter{equation} 
   \eqno{(\theequation{\mathit{a-d}})}
\end{equation*}
where $\tau_{w} = \left.\nu\frac{\partial\overline{u}}{\partial z}\right\rvert_{0}$ is the shear stress at the lower boundary. The respective exchanges of buoyancy and momentum  between the inner and outer layer are:
\begin{equation*}
\begin{aligned}
f_{um} &= -\overline{w'b'}|_{z_{um}} +\overline{b}(z_{um})\frac{\d z_{um}}{\d t}+\left. \kappa\frac{\partial \overline{b}}{\partial z}\right|_{z_{um}},\\
m_{um} &= \overline{w'u'}|_{z_{um}}-\overline{u}(z_{um})\frac{\d z_{um}}{\d t}. 
\end{aligned}
\label{eq:buomomflux} 
\refstepcounter{equation} 
\eqno{(\theequation{\mathit{a,b}})}
\end{equation*}
Here, $\overline{w'b'}|_{z_{um}}$ and $\overline{w'u'}|_{z_{um}}$ are the turbulent buoyancy and momentum fluxes at the level of the velocity maximum, $- \kappa({\partial \overline{b}}/{\partial z})|_{z_{um}}$ is the molecular buoyancy flux at the level of the velocity maximum and $\overline{b}(z_{um})\frac{dz_{um}}{dt}$ and $\overline{u}(z_{um})\frac{dz_{um}}{dt}$ are the Leibniz terms \citep{schatzmann1978leibniz, davidson1986discussleibniz, van2021unified} representing the effective buoyancy and momentum fluxes associated with a change in the height of the velocity maximum.

\begin{figure}
  \centerline{\includegraphics[scale =0.38, trim=0cm 0.2cm 13cm 0.8cm, clip]{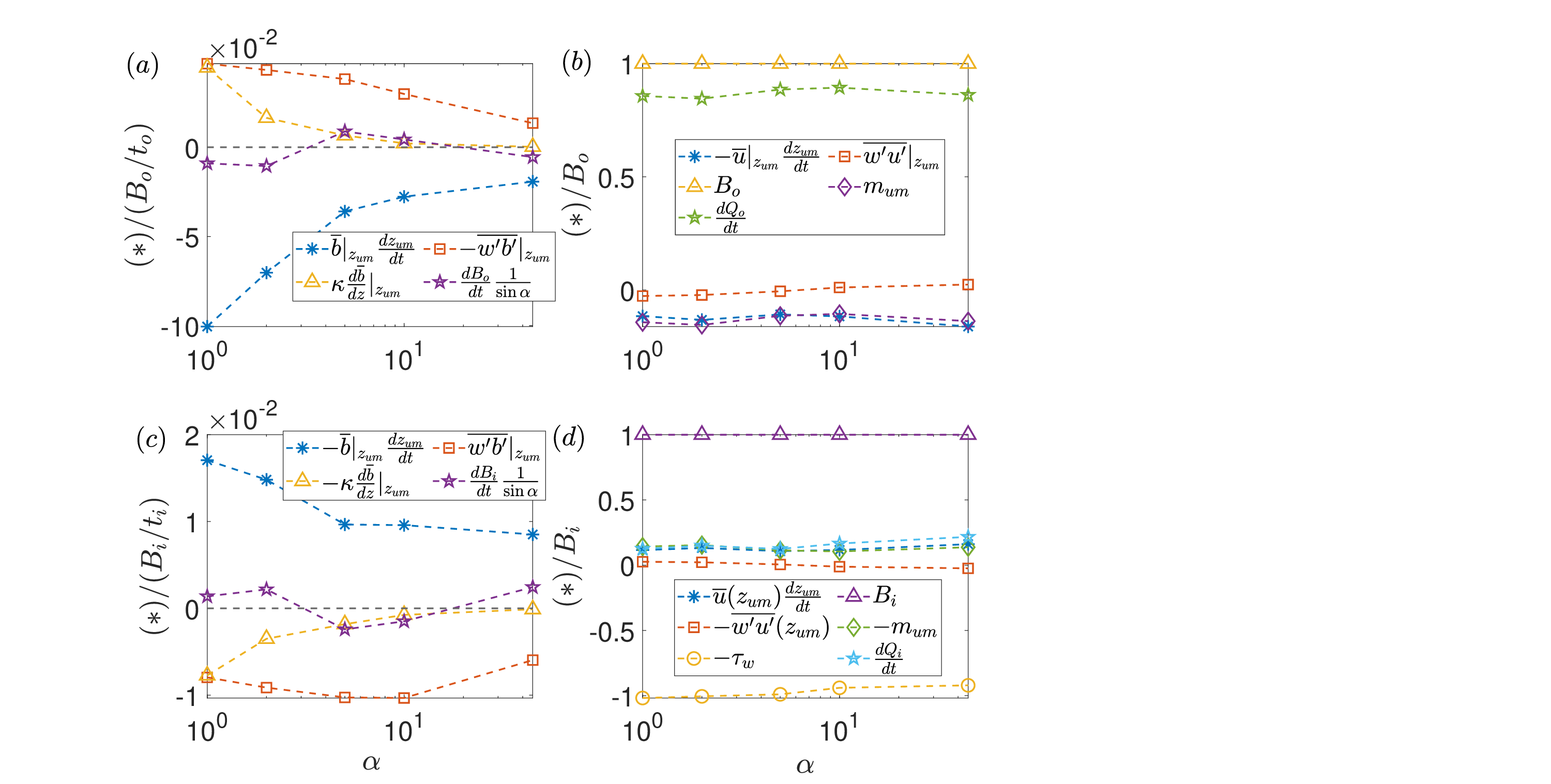}}
 \caption{Normalized terms against slope angles for the budgets of $(a)$ outer integral buoyancy forcing and $(b)$ outer volume flux , $(c)$ inner integral buoyancy forcing and $(d)$ inner volume flux. Note that these term are averaged over $t_{ave}$ in the dynamically equilibrium regime.}
\label{fig:Interaction_int}
\end{figure}

Figure \ref{fig:Interaction_int} ($a$) plots the terms in the integral buoyancy forcing budget of the outer layer (\ref{eq:IntMomBuoBud} $a$ and \ref{eq:buomomflux} $a$, scaled by ${B_o}/t_o$) as a function of slope angle, where $t_o= h_o/u_{To}$ is a typical turnover timescale of the outer layer. The magnitudes of the normalized fluxes are of order 10$^{-2}$, suggesting that  the exchange of buoyancy happens over timescales much longer than $t_o$. The flux with the largest magnitude is the Leibniz term (especially for small angles), but interestingly, the turbulent and molecular terms counteract it, creating a net buoyancy flux $f_{um}$ that is practically zero for all currents under consideration. 

Similarly, figure \ref{fig:Interaction_int} ($c$) shows the buoyancy flux terms from the inner layer budget (of equal magnitude and opposite sign to the outer layer budget), but instead normalized  by $B_i/t_i$, where $t_i= h_i/u_{Ti}$ is a turnover timescale of the inner layer. Here, the scaled budget terms are also of order 10$^{-2}$ and, as in the outer layer, the Leibniz term is in approximate balance with the turbulent and molecular terms. It is apparent that the integral buoyancy forcing $B_*$ can be approximated as constant in each of the inner and outer layers on timescales up to at least $t_i$ and $t_o$, respectively.

Given that the evolution of the integral buoyancy forcing ($dB_*/dt$) is not a leading order term in the inner and outer layer budgets at any slope angle considered, the dynamics governing the interface can therefore be regarded as quasi-steady in the self-similar regime and (\ref{eq:IntMomBuoBud}$a,b$) can be approximated as
\begin{equation}
     \frac{\d B_o}{\d t}\approx\frac{\d B_i}{\d t}\approx0,  \label{eq:B_intera}
\end{equation}
which is also consistent with the results in figure \ref{fig:evolution} $(j)$.

Equations (\ref{eq:IntMomBuoBud} $c,d$) and (\ref{eq:buomomflux} $b$) present the integral momentum (volume flux) budgets, where $B_*$ acts to accelerate the flow and increase the volume flux. Figure \ref{fig:Interaction_int} ($b$) illustrates the individual terms in the integral momentum budget of the outer layer, normalized by $B_o$.  At all angles it is clear that the Leibniz term makes the dominant contribution to the momentum exchange between layers $m_{um}$ (see \ref{eq:buomomflux} $b$), while the Reynolds stress $\overline{w'u'}|_{z_{um}}$ plays a negligible role. However, this momentum exchange has a magnitude of approximately 0.15$B_o$ at all angles considered and, therefore, plays a minor role in modifying the rate of volume flux increase, i.e.\ 
\begin{equation}
     \frac{\d Q_o}{\d t}\approx B_o.
\label{eq:Q_intera}
\end{equation}
Figure \ref{fig:Interaction_int} $(d)$ shows the integral momentum budget for the inner layer and we observe a leading order balance between the buoyancy forcing and bottom shear stress, i.e.\
\begin{equation}
 B_i\approx\tau_w.   
\label{eq:Bi_tauw}
\end{equation} 
The rate of increase of volume flux in the inner layer is of a similar order to the Leibniz term (towards which the Reynolds stress contribution is negligible, as in the outer layer).

\subsection{Layer-specific force balance}
\cite{turner1959} showed that gravity currents reach an equilibrium state where the gravitational forces are balanced by ``entrainment drag" and bottom friction. It is useful to interpret this finding in the light of weak interaction between the inner and outer layers. 

The starting point is the integral momentum balance for the entire layer, which can be obtained by adding (\ref{eq:IntMomBuoBud} $c$) and (\ref{eq:IntMomBuoBud} $d$) to give the time derivative of $Q(=u_Th)$
\begin{equation}
    \frac{\d Q}{\d t}=h\frac{\d u_T}{\d t}+u_T\frac{\d h}{\d t}= (B_o + B_i)-\tau_w.
    \label{eq:dQdt}
\end{equation}
In terms of top-hat variables \citep{turner1959}, this equation can be written as
\begin{equation}
h\frac{\d u_T}{\d t}=(B_o + B_i)-Eu_T^2-\tau_w,
    \label{eq:intForceBalance}
\end{equation}
where $E=u_T^{-1} \d h/\d t$ is the entrainment coefficient of a temporal gravity current \citep{van2018smallscale, van2019}. Since $u_T$ is expected to be constant in the dynamically equilibrated regime \citep{turner1959} (also see Figure \ref{fig:evolution}), (\ref{eq:intForceBalance}) simplifies to
\begin{equation}
    B_o+B_i=\tau_w+Eu_T^2.
    \label{eq:DecomintForceBalanceFinal}
\end{equation}
Given (\ref{eq:Bi_tauw}) and that $m_{um}$ is of a similar magnitude to both $(B_i - \tau_w)$ and $(dQ_o/dt - B_o)$, as observed in figures \ref{fig:Interaction_int} $(b)$ and $(d)$, we deduce that
\begin{equation}
   B_o\approx Eu_T^2.
   \label{eq:BoEut2}
\end{equation}

These results support distinct dynamics in the inner and outer layers. Buoyancy in the inner layer primarily overcomes the bottom friction, as shown in figure \ref{fig:dynamics} $(b)$, whilst the buoyancy in the outer layer overcomes drag associated with entrainment of ambient fluid. In the absence of significant exchange of momentum, the inner and outer layers are only weakly coupled (subject to the continuity condition at the velocity maximum). 

Importantly, our theoretical parameterization and DNS results show that this weak coupling is not confined to currents on small-angle slopes (with relatively strong stabilizing stratification), as conjectured by \cite{salinas2021nature}, but also applies on larger angle slopes (where the stabilizing stratification is relatively weak). This observation signifies that the weak coupling is not a result of a density interface forming near the velocity maximum. Instead, it appears to be a natural behaviour of inclined gravity currents.

Although it has been reported that there is a mismatch between the levels of the velocity maximum and zero turbulent shear stress \citep{salinas2021nature, wei2021layeredjet}, our results indicate that this is not a leading-order effect and the gradient-diffusion hypothesis remains useful, i.e.\ both the viscous shear stress and turbulent shear stress are approximately zero at the level of the velocity maximum, across which there is essentially no momentum exchange by turbulence or diffusion. This `decoupling' of the two layers explains why the outer layer behaves independently of bottom friction and much like a current on a free-slip slope described by \citet{van2019}.

\section{Inner-shear-layer scaling} \label{sec:inner}
\subsection{Inner layer profiles} \label{sec:innerProfiles}

\begin{figure}
  \centerline{\includegraphics[scale =0.29, trim=0.2cm 0cm 1cm 0.4cm, clip]{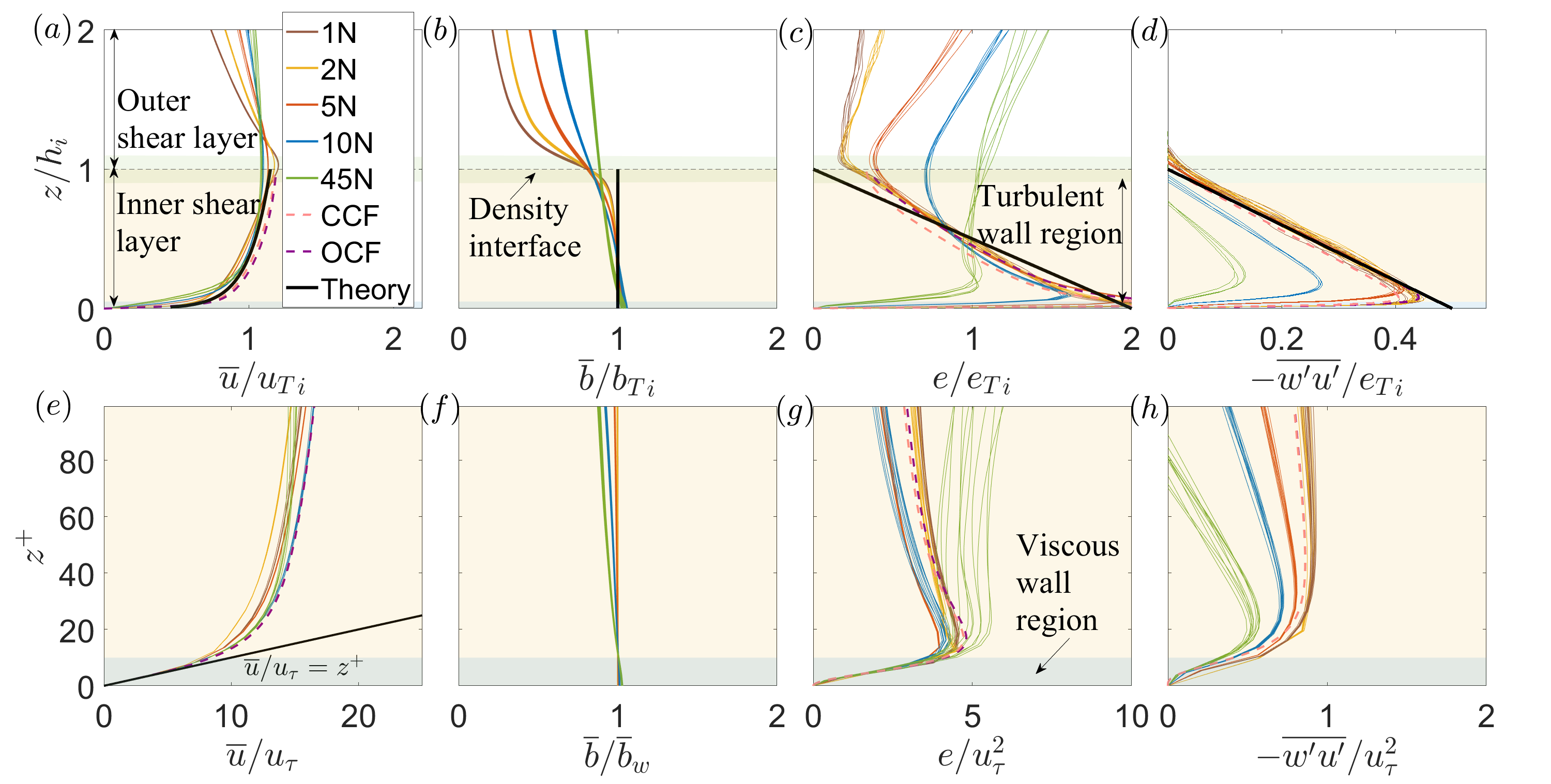}}
  \centerline{\includegraphics[scale =0.4, trim=0cm 0.5cm 0cm 0cm, clip]{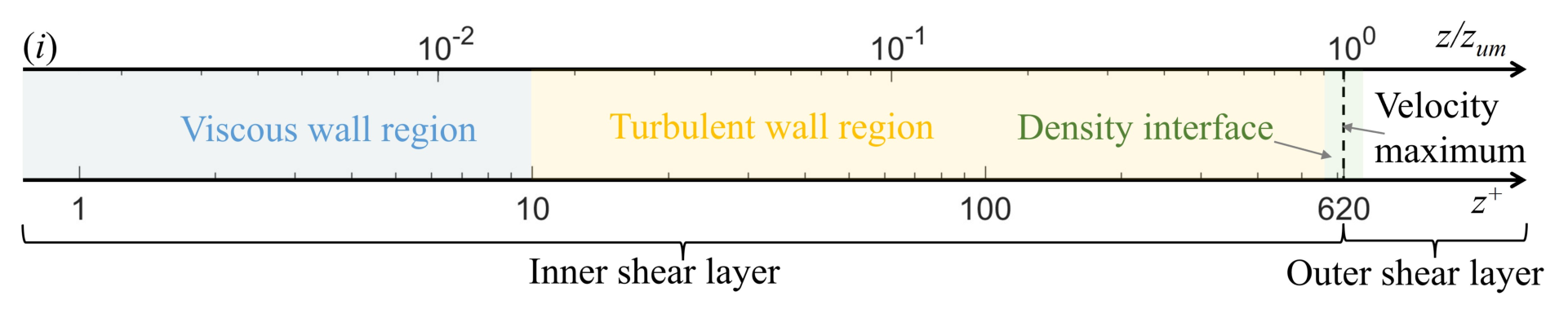}}
 \caption{Profiles of spatially averaged $(a)$ $\overline{u}/{u_T}_i$, $(b)$ $\overline{b}/{b_T}_i$, $(c)$ $e/{e_T}_i$ and $(d)$ $\overline{w'u'}/{e_T}_i$ against scaled height $z/h_i$; $(e)$ $\overline{u}/{u_\tau}$, $(f)$ $\overline{b}/\overline{b}_w$, $(g)$ $e/u_\tau^2$ and $(h)$ $\overline{w'u'}/u_\tau^2$ against $z^+ = zu_\tau/\nu$. Here, $u_\tau$ is the friction velocity, $\overline{b}_w$ is the spatially-averaged buoyancy at the wall. Note that the profiles at a series of times in the dynamically equilibrated regime are plotted for each case. CCF is the closed channel flow data, adapted from \cite{CCFlee2015}, and OCF is the open channel flow data adapted from \cite{OCFYao2022} both at $Re_\tau=550$. Distinct regions are highlighted with shading, which are depicted in panel $(i)$ using case 2N with $Re_\tau=620$, including the viscous wall region, the turbulent wall region and the density interface, represented by blue,yellow and green, respectively.} 
\label{fig:innerProfiles}
\end{figure}

Figure \ref{fig:innerProfiles} $(a-d)$ shows the inner layer profiles of $\overline{u}, \overline{b}, e$ and $\overline{w'u'}$ normalized by the characteristic scales  ${u_T}_i$, ${b_T}_i$ and ${e_T}_i$, respectively, at a series of times in the dynamically equilibrated regime. We employ $z/h_i$ as the scaled slope-normal coordinate. 

We decompose the inner shear layer into two regions: a turbulent wall region (TWR) and a viscous wall region (VWR) defined in terms of the dimensionless wall distance $z^+ = zu_\tau/\nu$. Figure \ref{fig:innerProfiles}$(i)$ illustrates these regions using case 2N. The VWR is defined up to $z^+ = 10$, rather than $z^+ = 50$ as in traditional boundary layers \citep{pope2000turbulent}, since the viscous contribution to total shear stress becomes negligible beyond $z^+ = 10$ for small-angled gravity currents. The TWR lies above the VWR, with TKE peaking at its lower boundary and decreasing to a minimum at its upper boundary. The upper boundary of TWR is defined via a density interface (see figure \ref{fig:innerProfiles} $(b)$) in the vicinity of the velocity maximum for small-angled cases. The distinct regions for case 2N are also illustrated in figure \ref{fig:innerProfiles} $(a-h)$ with the same colour scheme as in panel $(i)$. Note that the VWR occupies only a minor area at the bottom of the inner shear layer in panels $(a-d)$, as the vertical coordinate is scaled as $z/h_i$.

The profiles of scaled mean velocity in figure \ref{fig:innerProfiles} $(a)$ exhibit considerable collapse over a range of times. However, this  collapse is not indicative of a self-similar regime like the outer layer; instead, the inner layer reaches a quasi-steady state that evolves over very long timescales. Indeed, figure \ref{fig:dynamics} $(c)$ depicts (for a slope angle of 1$^\circ$) profiles of mean streamwise velocity $\overline{u}$ scaled by the initial (constant) velocity $u_0$ in the inner shear layer at different times. The profiles remain nearly unchanged with time because the buoyancy and shear stress gradient are in local balance and $\partial \overline{u}/\partial t \approx 0$ throughout the inner layer, as shown in figure \ref{fig:dynamics} $(b)$, i.e.\
\begin{equation}
  \frac{\d}{\d z} \left( \overline{w'u'}-\nu \frac{\partial \overline u}{\partial z} \right) \approx -\overline{b}\sin\alpha.
\end{equation}
Furthermore, the profiles of buoyancy in the inner layer (figure \ref{fig:innerProfiles} $b$) are practically uniform due to strong turbulent mixing, especially for small angles, implying that
\begin{equation}
    \overline{w'u'}-\nu \frac {\partial \overline u}{\partial z} \approx -\overline{b}z - \tau_w \approx B_i(z/h_i-1).
    \label{eq:totalshear}
\end{equation}
The approximately linear profile of total momentum flux suggested by \eqref{eq:totalshear} shares a clear analogy with a canonical plane turbulent channel flow subject to constant streamwise pressure gradient, where the momentum balance is given by ${\d} \left( \overline{w'u'}-\nu {\partial \overline u}/{\partial z} \right)/{\d z} =\d \overline{p}/dx.$

In order to explore this connection, profiles corresponding to a closed channel flow \citep[CCF, adapted from][]{CCFlee2015} and an open channel flow \citep[OCF, adapted from][] {OCFYao2022} with turbulent Reynolds numbers $Re_\tau=550$ have been included in figure \ref{fig:innerProfiles}. Note that CCF is subject to no-slip conditions on both the bottom and top boundaries of the channel, with symmetry about the plane at half height. For the purposes of comparison, the channel half-height is plotted as corresponding to $z/h_i =1$.  In contrast, OCF has a shear-free surface at the top, which is taken to correspond to $z/h_i =1$. The values of $Re_\tau$ that characterize the inner shear layer in the gravity currents are a function of slope angle (see table \ref{tab:sim} for more details), however, case 2N corresponds to $Re_\tau=620$, similar to that of the selected channel flow comparisons. The scaled velocity profiles of the channel flows shown in figure \ref{fig:innerProfiles} $(a)$ nearly overlap with those for the inner layer in the gravity currents, suggesting the strong similarity of these two flow types.

Figure \ref{fig:innerProfiles} $(c)$ shows the TKE profiles normalized by $e_{Ti}$. For small-slope angle currents, these profiles nearly collapse and decrease approximately linearly with height within the TWR. Deviations become apparent in the TKE profiles between gravity currents on small- and large-angled slopes, the likely explanation for which is that the density interface near the velocity maximum strengthens as the slope angle decreases (figure \ref{fig:innerProfiles} $(b)$), acting to suppress turbulence. This explanation is supported by the near collapse of the normalized TKE profiles for the small-angle cases (1N, 2N, 5N) with that for OCF, in which no vertical transport of turbulence is possible at the free surface. The magnitude of normalized TKE for CCF is slightly smaller than for OCF in the turbulent wall region, an effect attributed to stronger ``very-large-scale motions'' (VLSMs) in OCFs \citep{VLSMkim1999,VLSMbalakumar2007, OCFYao2022}. 

Figure \ref{fig:innerProfiles} $(d)$ shows the normalized turbulent shear stress $\overline{w'u'}/e_{Ti}$, which varies linearly with height throughout most of the inner layer  for small-angled gravity currents and all channel flows. This arises because viscous shear stress is negligible compared to turbulent shear stress in the TWR, allowing \eqref{eq:totalshear} to be simplified as
\begin{equation}
   \overline{w'u'} \approx B_i(z/h_i-1).
    \label{eq:wu_solution}
\end{equation}  
The observed collapse of these profiles upon scaling with $e_{Ti}$ is discussed in \S\ref{sec:steadySolution}.

Figure \ref{fig:innerProfiles} $(e-h)$ shows the scaled quantities in wall units with a focus on the VWR, which is highlighted using the same color scheme as panel $(i)$. The velocity profiles in Figure \ref{fig:innerProfiles}$(e)$ fully collapse in the VWR ($z^+ < 10$) across all cases, conforming to the well-known relationship $\overline{u}/u_{\tau} = z^+$ in this viscosity-dominated region. Figure \ref{fig:innerProfiles}$(f)$ shows that the  buoyancy throughout the VWR is close to the wall buoyancy, $\overline{b}_w$.

The profiles of TKE and $\overline{w'u'}$ normalized by $u_\tau^2$ are presented in Figure \ref{fig:innerProfiles}$(g)$ and $(h)$, respectively, in terms of wall units. The normalized TKE profiles are seen to collapse onto a single curve within the VWR that  increases rapidly  with height from zero on the lower boundary to a local maximum at the top of the VWR. The normalized $\overline{w'u'}$ profiles for the 2N slope current and the CCF and OCF cases, which all have a  similar $Re_\tau$, also nearly collapse in the VWR. However, the deviations from this normalized profile for the other slope currents considered suggests a possible $Re_\tau$ dependence in this scaling.



\subsection{Approximate steady-state solution}\label{sec:steadySolution}
Upon inspection, we find that an approach analogous to that for the outer layer in \S\S\ref{sec:outerSolutions} and \ref{sec:paraTKE} can also be applied to the inner shear layer. On the basis of the observed collapse in figure \ref{fig:innerProfiles}, approximate solutions are proposed, especially for currents on small-angled slopes:  \begin{equation*}
 \overline{u}={u_T}_i {f_u}_i(\zeta), \quad \overline{b}={b_T}_i {{f_b}_i}(\zeta),\quad e={e_T}_i {{f_e}_i}(\zeta), \quad \zeta=z/h_i \in (0, 1).
  \refstepcounter{equation} 
  \eqno{(\theequation{\mathit{a-c}})}
  \label{eq:basicInnerfufb}
 \end{equation*}
 Note that $f_{ui}(\zeta)$, $f_{bi}(\zeta)$, and $f_{ei}(\zeta)$ are expected to be strictly valid only within the TWR. However, as the VWR volume is negligible compared with that of the TWR for the small-angled slope currents, we assume that these functions may be applied throughout the inner shear layer. Taking $f_{bi}(\zeta)\approx 1$ (consistent with figure \ref{fig:innerProfiles}$(b)$), we assume that ${f_u}_i$ takes a logarithmic form, as for an unstratified boundary layer adjacent to a no-slip surface:
\begin{equation}
   {f_u}_i(\zeta)= c_i\ln{\zeta}+c_{um},
   \label{eq:fuifinal}
\end{equation}
where $c_i$ is a coefficient to be determined and $c_{um}={f_u}_i(1)=\overline{u}_m/{u_T}_i$.
Consistency with the volume transport decomposition for the inner layer in (\ref{eq:QMB} $a$) requires
\begin{equation}
    Q_i=\int_0^{h_i}\overline{u}dz = {u_T}_i h_i\int_{0}^1{f_u}_i(\zeta)d\zeta={u_T}_ih_i
    \label{eq:integralu} \quad\Rightarrow\quad \int_{0}^1{f_u}_i(\zeta)d\zeta=1,
\end{equation}
and thus $c_i=c_{um}-1$.

As in \S\ref{sec:paraTKE}, we propose that $e$ and $\overline{w'u'}$ in the inner layer (where $\partial \overline{u}/\partial z$ is now $> 0$) are related by the scaling
\begin{equation}
    e=K_{mi}S/c_{mi}=-\overline{w'u'}/c_{mi}.
   \label{eq:inner_ecm}
\end{equation}
 Note that an additional subscript $i$ is used to distinguish the inner shear layer eddy parameterization coefficient ($c_{mi}$ and, later, $c_{\rho i}$ and $c_{\epsilon i}$) from that applicable to the outer shear layer. Combining  \eqref{eq:wu_solution}, \eqref{eq:basicInnerfufb} and \eqref{eq:inner_ecm} suggests
that $f_{ei}$ takes a linear form (consistent with collapse of the small-angled cases in figure \ref{fig:innerProfiles}$(c)$) and consistency with the TKE decomposition for the inner layer \eqref{eq:e_T} requires that
$\int_{0}^{1}f_{ei}(\zeta)d\zeta=1$, thus
\begin{equation}
    f_{ei}=2(1-\zeta), \quad 
e_{Ti}=e/f_{ei}=B_i/(2c_{mi}).
    \label{eq:innereToBi}
\end{equation}
The approximate solutions proposed in \eqref{eq:basicInnerfufb} and \eqref{eq:inner_ecm} can thus be summarized as
\begin{equation*}
   \frac{\overline{u}}{{u_T}_i}= \underbrace{(c_{um}-1)\ln{\zeta}+c_{um}}_{{f_u}_i(\zeta)}, \quad \frac{\overline{b}}{b_{Ti}}=\underbrace{1}_{f_{bi}(\zeta)},\quad \frac{e}{e_{Ti}} = -c_{mi}\frac{\overline{w'u'}}{e_{Ti}}=\underbrace{2(1-\zeta)}_{f_{ei}(\zeta)}.
   \refstepcounter{equation} 
  \eqno{(\theequation{a-c})}
  \label{eq:innerfube} 
\end{equation*}

The ratio 
 of $u_m$ to $u_{Ti}$ (i.e.\  $c_{um}$) 
is found from the DNS data to be approximately 1.12. It is interesting to consider the analogy with channel flow using the well-known approximation for the inner layer, $\overline{u}_m =c_\kappa^{-1} {\ln z^+} +C$ (\cite{pope2000turbulent}; equations (7.43 and 7.44)),
where $c_\kappa=0.41$ is the von Kármán constant and $C$ is a constant with a value of about 5.2. Comparison with (\ref{eq:innerfube}$a$) suggests that
\begin{equation}
c_{um} = \frac{c_{\kappa}C + \ln Re_{\tau}}{c_{\kappa}C + \ln Re_{\tau}-1}, \quad
\frac{u_{Ti}}{u_\tau} =\frac{1}{(c_{um}-1)c_\kappa},  
\label{eq:log_analogy}
\end{equation}
which is in excellent agreement with the DNS data, e.g.\ yielding $c_{um} = 1.13$ and $u_{Ti}/u_{\tau} = 18.8$ for $Re_{\tau} = 620$. 

\begin{figure}
   \centerline{\includegraphics[scale =0.3, trim=0cm 12.5cm 2.5cm 1.2cm, clip]{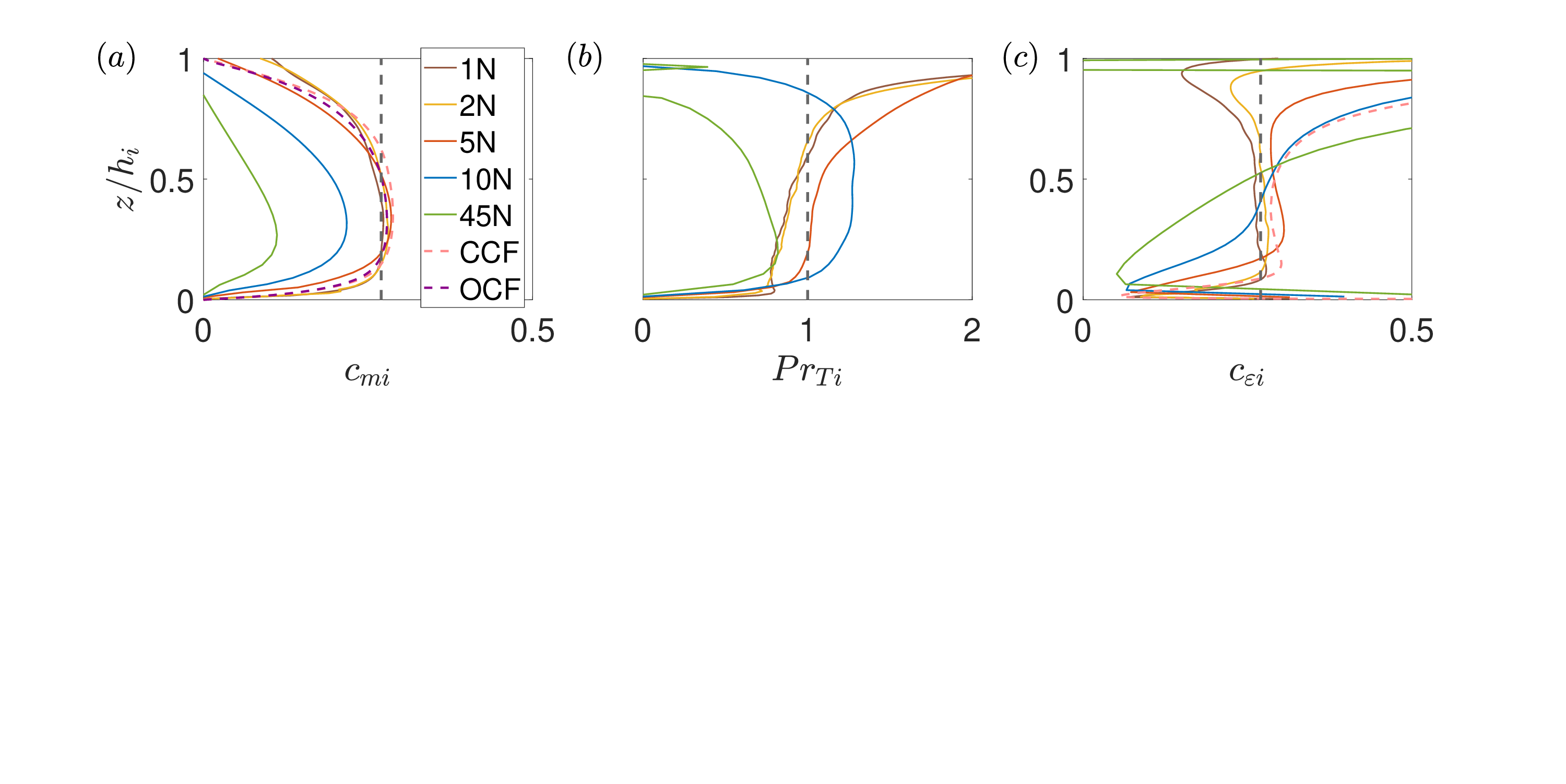}}
  \caption{ Scaling of inner layer turbulence parameters averaged over $t_{ave}$ in the dynamically equilibrated regime: $(a)$ $ c_{mi}=K_{mi}S/e\approx 0.27$, $(b)$ $ Pr_{Ti}=c_{mi}/c_{\rho i}\approx1$ and $(c)$ $ c_{\varepsilon i}=\varepsilon/(eS)\approx0.27$  against $z/h_i$. The converged values are denoted with the vertical dashed lines.} 
 \label{fig:TurParInner}
\end{figure}

Figure \ref{fig:TurParInner} $(a), (b)$, and $(c)$ show the coefficients $c_{mi}$, $Pr_{Ti}$ and $c_{\varepsilon i}$ based on the DNS data and the form of the outer layer scaling in \eqref{eq:shearScaling}. We observe that a single value for each scaling coefficient can be applied with reasonable success to the core region of the TWR (where $z/ h_i \in [0.25, 0.75]$) in the small-angle slope currents (1N, 2N, 5N) and the channel flows. 
In particular, $c_{mi}$ and $c_{\varepsilon i}$ both converge to approximately 0.27 in the core region, while $Pr_{Ti}$ is close to unity, consistent with the well-known Reynolds analogy. 
The functions given by \eqref{eq:innerfube} with $c_{mi} = 0.27$ and $c_{um} = 1.12$ are plotted in figure \ref{fig:innerProfiles} $(a-d)$, and show reasonable agreement with the DNS data for currents on small-angle slopes (e.g.\ up to 5$^{\circ}$) within the TWR. The corresponding prediction from \eqref{eq:innereToBi} that  $e_{Ti}/B_i\approx1.9$  is plotted in figure \ref{fig:averagedVars} $(c)$ and agrees well with the DNS results for small-angled currents. 
 We suggest that the likely reason for the success of this scaling is that the core region of the TWR is sufficiently far from both the wall and the density interface for the turbulence to be shear-dominated \citep{mater2014threeregime} and, therefore, that parameterizing the turbulence using $|\partial \overline{u}/\partial z|$ and $e$ is reasonable. 

\section{Inner-outer-layer matching condition}\label{sec:matchCond}
\subsection{Buoyancy partition}
\label{sec:buoyPart}
\begin{figure}
  \centerline{\includegraphics[scale =0.32, trim=0.8cm 13.8cm 0.8cm 1.5cm, clip]{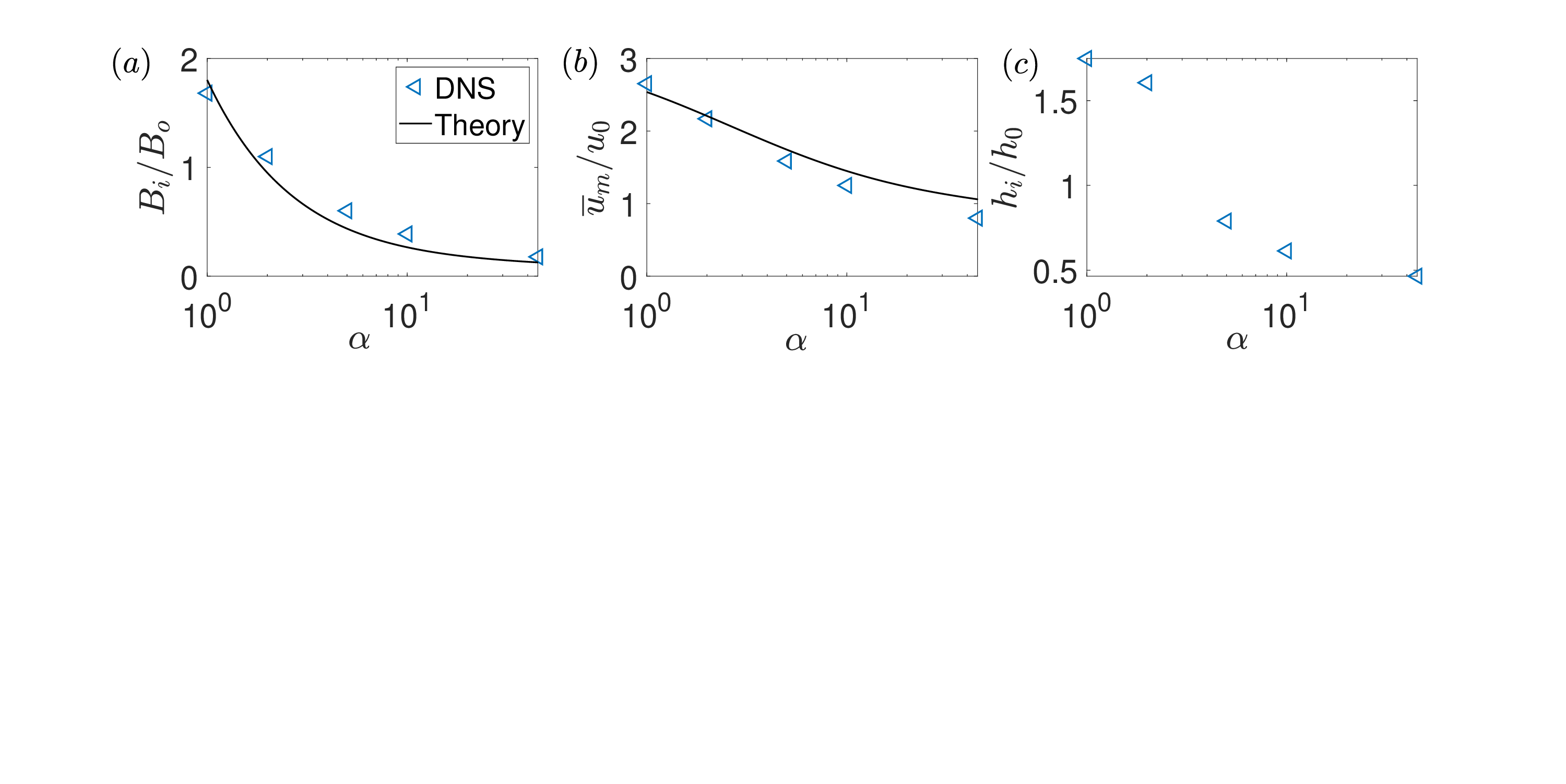}}
  \caption{$(a) B_i/B_o$, $(b) \overline{u}/u_0$ and $(c) h_i/h_0$ averaged over $t_{ave}$ in the dynamically-equilibrated regime as a function of slope angle $\alpha$ (degrees). The solid lines in panels $(a-b)$ denote the theoretical prediction in \eqref{eq:BiOBo_final}, \eqref{eq:umSolu}, respectively. }
\label{fig:BuoyPart}
\end{figure}

It is clear that the buoyancies in the inner and outer layer ($B_i$ and $B_o$) are the primary forcing in that layer. Therefore, a crucial step in obtaining a closed-form description of a slope current is to predict how buoyancy is partitioned between the layers. Given that $u_T\approx u_{To}$ (figure \ref{fig:evolution} g) and $\tau_w=u_\tau^2$, \eqref{eq:Bi_tauw} and \eqref{eq:BoEut2} can be rewritten as:
\begin{equation*}
    B_o=Eu_{To}^2,\quad\quad B_i=u_\tau^2.
    \refstepcounter{equation}
    \eqno{(\theequation{a,b})}
    \label{eq:Burelate}
\end{equation*}
Combining this with a velocity-matching condition in terms of the self-similar relations $\overline{u}_m=3u_{To}/2=c_{um}u_{Ti}$ gives
\begin{equation*}
    B_o=\frac{4}{9}E\overline{u}_m^2 ,\quad B_i=c_{mk}^2\overline{u}_m^2, \quad{\rm and \, thus \,\,}
    \frac{B_i}{B_o}=\frac{9}{4}c_{mk}^2E^{-1},
     \refstepcounter{equation}
    \eqno{(\theequation{a-c})}
    \label{eq:BiOBo}
\end{equation*}
where $c_{mk}={(c_{um}-1)c_{\kappa}}/{c_{um}}$.
Recalling from 
 (\ref{eq:fufbfe} $a$) that $E=B_o/u_{To}^2=a_u^{-2}$, where $a_u$ is given by (\ref{eq:mvrSolu} $a$), the predicted dependence of $B_i/B_o$ on $\alpha$ is 
\begin{equation}
    \frac{B_i}{B_o}=\frac{2c_{mk}^2(Pr_T+{c_m}/\tan\alpha)}{Pr_T(c_m-c_\varepsilon) },
    \label{eq:BiOBo_final}
\end{equation}
which is compared in Figure \ref{fig:BuoyPart} $(a)$  with $B_i/B_o$ calculated from the DNS data. Reasonably good agreement is found across the range of slope angles $\alpha$ considered and the ratio $B_i/B_o$ is seen to increase as both the slope angle $\alpha$ and the associated entrainment rate  decrease (equation \ref{eq:BiOBo} $c$). This insight could explain the long-runout of submarine gravity currents over mild slopes. As the slope angle reduces: 1) a greater proportion of the buoyancy is confined in the inner layer where it propagates largely undiluted because of a weak interaction with the outer layer; and 2)  the outer-layer buoyancy also experiences limited dilution due to the reduced entrainment rate.

 The maximum velocity $\overline{u}_m$ is determined by substituting \eqref{eq:BiOBo_final} and $B=B_i+B_o$ into \eqref{eq:BiOBo} to give

\begin{equation}
\overline{u}_m(\alpha)=\frac{\sqrt{B_i}}{c_{mk}}=\sqrt{\frac{2B( Pr_T+c_m/\tan\alpha)}{2c_{mk}^2(Pr_T+c_m/\tan\alpha)+Pr_T(c_m-c_\varepsilon)}}=\frac{3}{2}u_{To}=c_{um}u_{Ti}.
    \label{eq:umSolu}
\end{equation}
The DNS data for $\overline{u}_m$ and $u_{T*}$ shown in figure \ref{fig:BuoyPart} $(b)$ and figure \ref{fig:averagedVars} $(a)$ are seen to be well-predicted by \eqref{eq:umSolu}. Notably, the characteristic thickness $h_i$ remains a free parameter. Applying buoyancy conservation yields
\begin{equation}
  \frac{\int_0^{z_{um}}\overline{b}(t,z)\d z}{\int_0^\infty \overline{b}(t=0,z)\d z}=\frac{b_{Ti}h_i}{b_0h_0}=\frac{B_i}{B_o+B_i}\quad\Rightarrow\quad\frac{h_i}{h_0} = \frac{B_i/B_o}{1+B_i/B_o}\frac{b_0}{b_{Ti}}.
    \label{eq:hisolu}
\end{equation}
Here, $B_i/B_o$ is a function of $\alpha$ as given in \eqref{eq:BiOBo_final}. \eqref{eq:hisolu} indicates $h_i/h_0$ ($h_0$ set to constant) depends on $\alpha$ and the ratio of the initial buoyancy $b_0$ to the inner characteristic buoyancy $b_{Ti}$. As shown in figure \ref{fig:BuoyPart} $(c)$, $h_i/h_0$ attains larger values at smaller angles, with a sharp increase observed between cases 2N and 5N. This sudden rise is likely driven by the dilution due to the initial burst (see $t/t^*\in[20, 40]$ in figure \ref{fig:evolution}), which substantially increases the ratio $b_0/b_{Ti}$. This observation somewhat suggests that the history of a flow influences its subsequent evolution, as highlighted by \cite{caulfield2021layering}. 

\subsection{Entrainment law}
\label{sec:mixEntrain}
Using the self-similar relations for $\overline{u}_m$, $u_{To}$ and $u_{Ti}$ in \S\ref{sec:buoyPart}, we can obtain expressions for $Ri_*$ upon combining (\ref{eq:hubRi} $d$), \eqref{eq:BiOBo_final}, \eqref{eq:umSolu} and $u_T \approx u_{To}$:

\begin{equation*}   
    Ri_o=\frac{9}{8}\frac{Pr_T(c_m-c_\varepsilon)}{\tan\alpha Pr_T +c_m},\quad
    Ri_i=\frac{c_{mk}^2c_{um}^2}{\tan\alpha},\quad Ri=\frac{9}{8}\frac{2c_{mk}^2(Pr_T+c_m/\tan\alpha)+Pr_T(c_m-c_\varepsilon)}{\tan\alpha Pr_T+c_m}.
    \refstepcounter{equation}
    \eqno{(\theequation{a-c})}
    \label{eq:Risolu} 
\end{equation*}
The predictions for these Richardson numbers are plotted in Figure \ref{fig:averagedVars} $(b)$ with good agreement apparent for small-slope angles. Equation (\ref{eq:Risolu} $a$) indicates the outer layer Richardson number $Ri_o$ increases and approaches a finite limit ($Ri_{om} \approx 9{Pr}_T (c_m-c_\epsilon)/8c_m$) as $\alpha$ decreases ($\tan\alpha \ll c_m/{Pr}_T$). This limiting value suggests that the outer layer remains marginally stable and weakly stratified at all the slope angles considered here. However, $Ri$ and $Ri_i$ continue to increase as the slope angle decreases. Notably, $Ri_i$ loses physical relevance for small-angled cases, as the inner layer is nearly well-mixed in these scenarios. $Ri$ is directly related to the buoyancy partition, as shown by the following relationship derived from (\ref{eq:hubRi} $d$) with $u_T \approx u_{To}$:
\begin{equation}
    Ri/Ri_o = B/B_o.
    \label{eq:RiORio}
\end{equation}
Therefore, the increase in $Ri$ with decreasing $\alpha$ essentially reflects the increasing proportion of the integral buoyancy held in the inner layer. 

\begin{figure}
  \centerline{\includegraphics[scale =0.37, trim=0.4cm 0.3cm 0.8cm 1.8cm, clip]{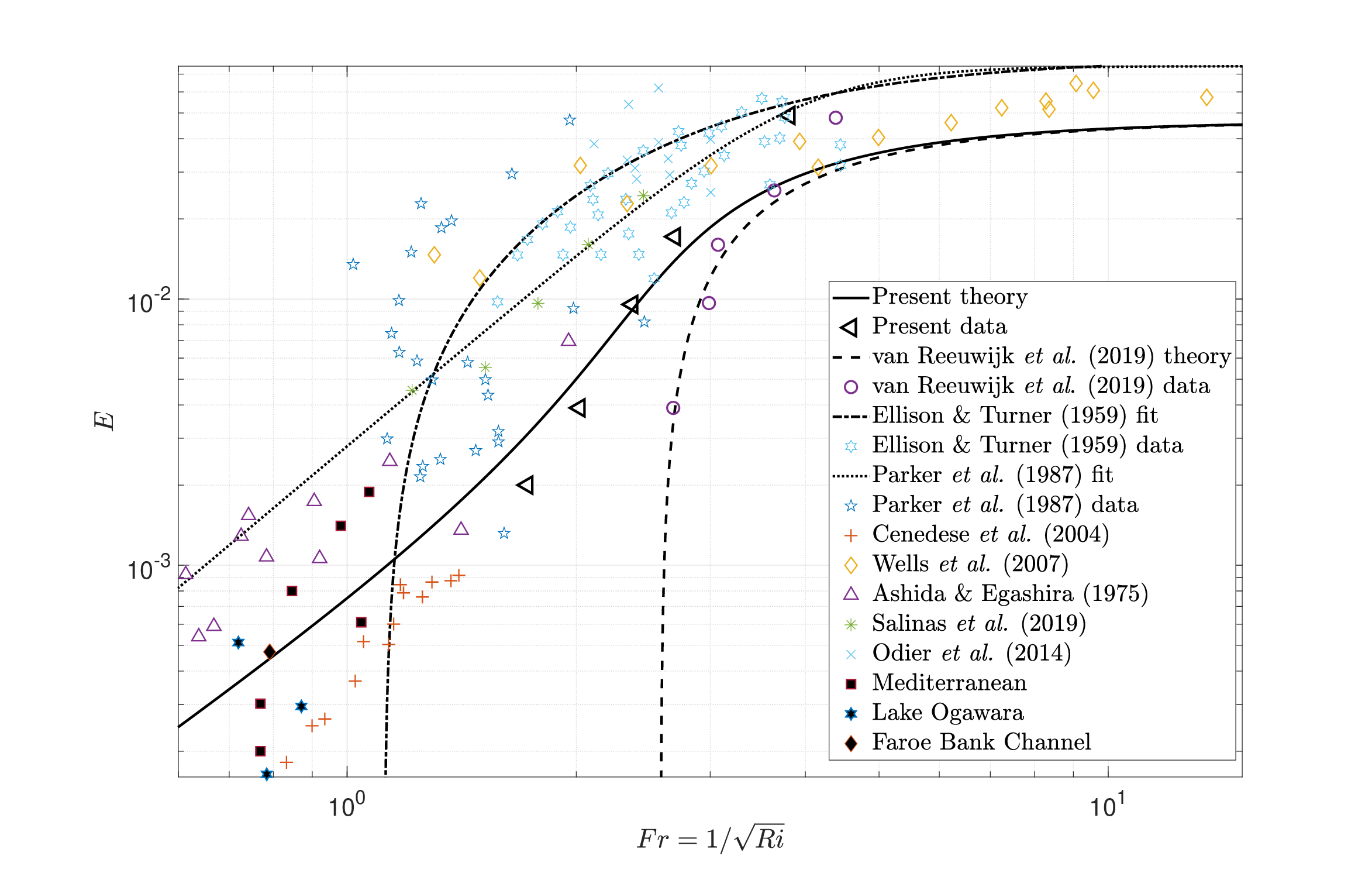}}
  \caption{Entrainment rate $E$ against $Fr=1/\sqrt{Ri}$. The black solid line and triangles denote the prediction in \eqref{eq:EvsRi} and the present DNS data, respectively. The data and theory from \cite{van2019} and the previous data and fitted functions  compiled by  \cite{odier2014linearity} and \cite{salinas2019againstslope} are also shown  (incorporating the studies by \cite{turner1959}, \cite{parker1987experiments}, \cite{cenedese2004dense}, \cite{wells2007influence}, \cite{ashida1975basic},  \cite{odier2014linearity} and  \cite{salinas2019againstslope}, together with field data (filled markers) collected from the Mediterranean, Lake Ogawara, and the Faroe Bank Channel). Dataset for this plot will be made available upon publication.}
\label{fig:FrE}
\end{figure}

As we have shown in this study that  entrainment is associated with the outer layer dynamics in a slope current, 
we now adapt the entrainment law derived by \cite{van2019} for  currents on free-slip boundaries (equations (4.24) and (4.25) therein). Using \eqref{eq:RiORio} to relate $Ri$ and $Ri_o$,
their theoretical  expression in terms of $Ri_o$ can be rewritten as
\begin{equation}
    E=\frac{c_m}{Pr_T}(Ri_{om}-Ri_o)=\frac{c_m}{Pr_T}\left(Ri_{om}-Ri\frac{B_o}{B}\right),\quad Ri_{om}\approx0.15,
    \label{eq:EvsRi_ini}
\end{equation}
which, with (\ref{eq:BiOBo} $c$), gives the entrainment law as an explicit  function of $Ri$:
\begin{equation}
   E=\frac{c_m}{Pr_T}\left(Ri_{om}-\frac{ERi}{9c_{mk}^2/4+E}\right) ~\Rightarrow~  E=\sqrt{\left(\frac{c_mRi}{2Pr_T}+c_{R1}\right)^2+\frac{c_mc_{R2}}{Pr_T}}-\frac{c_mRi}{2Pr_T}-c_{R1},
    \label{eq:EvsRi}
\end{equation}
where $c_{R1}={9}c_{mk}^2/{8}-{c_m}Ri_{om}/({2Pr_T})$ and $ c_{R2}={9}c_{mk}^2{c_m}Ri_{om}/(4Pr_T)$ are constants. 

Figure \ref{fig:FrE}  shows the predicted entrainment rate $E$ as a function of densitometric Froude number $Fr=1/\sqrt{Ri}$. The present theory \eqref{eq:EvsRi} (solid black line) is shown together with entrainment models from \citet[][dashed line]{van2019}, \citet[][dash-dotted line]{turner1959}, and \citet[][dotted line]{parker1987experiments} along with a broad dataset (denoted by coloured symbols) from laboratory experiments, DNS and field observations, as compiled by \cite{odier2014linearity} and \cite{salinas2019againstslope} (details provided in the caption). 

The classical parametrisation based on the experimental data from \cite{turner1959} \citep[proposed by][] {turner1986assumption}  aligns well with the high-$Fr$ data, but shows a different asymptotic behaviour at  low-$Fr$ (with $E$ dropping to 0 at a critical $Fr$ value beyond the range accessible to their  experiments).
The parametrisation fitted by \cite{parker1987experiments} has  asymptotic behaviour that is more consistent with the data and offers better overall performance, but its functional form lacks a solid theoretical basis.

Comparison of the theoretical predictions of \cite{van2019} and the current study in figure \ref{fig:FrE} highlights the significant role that the outer layer dynamics is likely to play in many slope current applications. The two predictions show consistency with each other 
at large $Fr$ (corresponding to steep slopes), where the integral buoyancy forcing is confined primarily within the outer layer (i.e.\ $B_o/B\approx1$). Both theories suggest 
a maximum entrainment rate $E_m = c_mRi_{om}/Pr_T\approx0.046$ as $Fr$ approaches infinity, which is in good agreement with the value of 0.04 proposed by \cite{wells2010relationship}. 
As $Fr$ decreases, the present theory \eqref{eq:EvsRi} indicates that entrainment is not completely suppressed at a critical $Ri$ ($Fr$), but rather asymptotes towards 0 as $Ri\rightarrow\infty (\alpha\rightarrow0)$, consistent with the level of TKE in the outer layer $e_{To}$ that scales with $B_o$ (see \ref{eq:outereoB}). Notably, this differs conceptually from the hypothesis of `continued (high-Richardson-number) mixing' associated with intermittent turbulence under strong stratification \citep[see e.g.,][]{wells2010relationship}. Despite $Ri$ approaching infinity as $\alpha$ decreases towards 0, the outer layer herein remains weakly stratified with $Ri_o$ asymptotically approaching $Ri_{om}$ as discussed earlier. Crucially, the present theory shows good agreement with the field data (filled symbols in figure \ref{fig:FrE}), offering a physical basis and the prospect of general  applicability to flows at small $Fr$ of geophysical relevance.

\section{Conclusion}\label{sec:conclusion}
In this paper, we explored the fundamental flow structure and scaling laws of temporal inclined gravity currents using direct numerical simulations. The simulations run for a duration that is sufficient to reach a dynamically equilibrated (time-evolving) regime across a range of slope angles. We find that the slope currents comprise a relatively well-mixed inner layer adjacent to the slope that is overlain by a density-stratified outer layer. The inner and outer layers are delineated by the level at which a velocity maximum is situated. In the dynamically equilibrated regime, the outer layer exhibits self-similar dynamics identical to those of gravity currents on free-slip slopes studied by \cite{van2019}. The inner layer resembles fully developed plane turbulent channel flow, in which the shear stress decreases linearly with distance from the wall and the logarithmic velocity defect law applies. 

At small slope angles, a density interface is observed to form in the vicinity of the velocity maximum. Although the presence of a density interface has been interpreted in the literature as a decoupling between the inner and outer layers \citep{selfsharp2019, salinas2021nature}, our simulations indicate that the two layers are effectively decoupled for all slope angles investigated. As a consequence, the integral buoyancy and volume flux in each layer evolve nearly independently (subject to the continuity condition at the maximum). The classic force balance, in which buoyancy forces are countered by entrainment drag and wall friction, can be further refined: the outer layer buoyancy forcing is responsible for overcoming the entrainment drag, whilst the inner layer buoyancy forcing counteracts the wall friction. 

Based on the flow structure, we have developed a theoretical description of an inclined gravity current by matching the dynamics of a turbulent wall-bounded inner layer and a self-similar outer layer at the velocity maximum. The theory predicts 
the flow quantities as functions of slope angle only, and are expected to best characterize currents 
with higher friction Reynolds numbers $Re_\tau$ (corresponding to smaller slope angles in this study), for which the inner layer is more analogous to a pressure-driven channel flow and the core region of the layer is sufficiently far from both the wall and the density interface for the turbulence to be shear-dominated \citep{mater2014threeregime}.

An important observation in both the simulations and the theory is that the ratio of the integral buoyancies in the inner and outer layer increases as the slope angle decreases. This insight offers a potential explanation for the long-runout of submarine gravity currents along mild slopes: as the slope angle reduces, firstly, a greater proportion of the buoyancy is confined in the inner layer (where it remains  largely undiluted because of a weak interaction with the outer layer) and, secondly, entrainment of ambient fluid into the outer-layer (and consequent dilution of its buoyancy) is also reduced. The theory also gives the entrainment rate $E$ as a function of the overall Richardson number $Ri$. The entrainment model allows application to small slope angles of oceanographic relevance and aligns well with field data collected from the Mediterranean, Lake Ogawara, and the Faroe Bank Channel. Although the minimum slope angle considered in the simulations here is 1$^\circ$, the inner-outer scaling offers a solid physical basis from which the theoretical predictions have been extrapolated to the milder slopes that characterize a range of geophysical flows.



One interesting question this study poses is whether inclined gravity currents can reach a strongly stratified regime—specifically, whether they can enter the so-called (high Richardson number) ‘right flank’ \citep{linden1979mixing,wells2010relationship,caulfield2021layering}. Our results indicate that even though the bulk Richardson number $Ri$ can exceed $1/4$ (and approach infinity when $\alpha\rightarrow0$), a threshold often associated with ‘marginal stability’ \citep{thorpe2009marginal}, neither the inner layer nor the outer layer become strongly stratified. In contrast, the outer Richardson number $Ri_o$ remains below $1/4$ regardless of the slope angle and appears to be a more relevant measure of the dynamical importance of the stratification. This highlights the importance of the layer-wise perspective proposed in this paper for analysing complex geophysical wall-bounded stratified flows, where an (inner) unstratified boundary layer can coexist with an (outer) stratified shear layer. A bulk model for the current may overlook the key physics and internal processes governing the flow.

\backsection[Funding]{
The authors acknowledge the UK Turbulence Consortium (EPSRC grant EP/R029326/1 and EP/X035484/1), for the grand challenge project that provided the computational resources for this work. Lianzheng Cui acknowledges the financial support from China Scholarship Council for his PhD study.}

\backsection[Declaration of interests]{The authors report no conflict of interest.}

\backsection[Data availability statement]
{Supporting data for this paper will be made available upon publication.}


\bibliographystyle{jfm}
\bibliography{jfm}



\end{document}